\documentclass[12pt]{article}
\usepackage{amsfonts}
\usepackage{amsmath}
\usepackage{amssymb}
\usepackage{mathptmx}
\usepackage{geometry}
\usepackage{graphicx}
\usepackage{hyperref}
\usepackage{cancel}
\usepackage{textcomp}
\usepackage{tikz}
\usepackage{bm}
  \usepackage{mathrsfs}
  \usepackage{filecontents}
\usepackage{times}
\usepackage{epsfig}
\usepackage{xcolor}
\usepackage{slashed}
\usepackage{booktabs}
\usepackage{latexsym}
\usepackage{verbatim}
\usepackage{extarrows}
\usepackage{multirow}
\usepackage{rotating}
\usepackage{colortbl}
\usepackage{indentfirst}

\definecolor{mygray}{gray}{.9}
\definecolor{intnull}{RGB}{213,229,255}
\usepackage{arydshln}
\usepackage{diagbox}

\usepackage{caption}
\usepackage{tikz}
\usetikzlibrary{arrows,shapes,chains}
\usepackage{graphicx, subfig}
\begin{document}
\renewcommand{\thefootnote}{\fnsymbol{footnote}}
\baselineskip=16pt
\pagenumbering{arabic}
\vspace{1.0cm}
\begin{center}
{\Large\sf  Quasinormal mode and spectroscopy  of a Schwarzschild black hole surrounded by a cloud of strings in Rastall gravity}
\\[10pt]
\vspace{.5 cm}

{Xin-Chang Cai\footnote{E-mail address: caixc@mail.nankai.edu.cn} and Yan-Gang Miao\footnote{Corresponding author. E-mail address: miaoyg@nankai.edu.cn}}
\vspace{3mm}

{School of Physics, Nankai University, Tianjin 300071, China}

\vspace{4.0ex}

{\bf Abstract}
\end{center}

We obtain the solution of a static spherically symmetric black hole surrounded by a cloud of strings in Rastall gravity and study the influence of the parameter $a$ associated with strings on the event horizon and the Hawking temperature. Through analyzing the black hole metric, we find that the static spherically symmetric black hole solution surrounded by a cloud of strings in Rastall gravity can be transformed into the static spherically symmetric black hole solution surrounded by quintessence in Einstein gravity when the parameter $\beta$ of Rastall gravity is positive, which provides a possibility for the string fluid to be a candidate of dark energy. We use the method of Regge and Wheeler together with the high order WKB-Pad\'{e} approximation to calculate the quasinormal mode frequencies of the odd parity gravitational perturbation and simultaneously apply the unstable null geodesics of black holes to compute the quasinormal mode frequencies at the eikonal limit for this black hole model. The results show that the increase of the parameter $a$ makes the gravitational wave decay slowly in Rastall gravity. In addition, we utilize two methods, which are based on the adiabatic invariant integral and the periodic property of outgoing waves, respectively, to derive the area spectrum and entropy spectrum of the black hole model. The results show that the area spectrum and entropy spectrum are equidistant spaced. The former is same as the case of Einstein gravity, while the latter is different, depending on the Rastall parameter $\beta$.

\newpage

\section{Introduction}
The gravitational waves from binary black hole or neutron star mergers detected by LIGO and Virgo collaborations~\cite{P1,P2}, as well as the first black hole photograph released recently, have fueled our interest in black holes. A black hole is a magic celestial body predicted by Einstein's general relativity and is a spacetime region where the light cannot escape. An important means to obtain the basic properties of black holes is to study quasinormal modes in the spacetime of black holes. In the past few decades, the quasinormal modes of various classical or semiclassical black holes have been studied extensively and deeply under the framework of Einstein's theory of gravity. Up to now, the conservation law that the covariant  derivative of the energy-momentum tensor in Einstein gravity is zero has only been tested~\cite{YF} in the Minkowski spacetime or specifically in a weak gravitational field limit, so the covariant  derivative may not be zero in a strong gravitational field. Therefore, it is especially necessary for the analysis of gravitational waves to study the quasinormal modes of black holes in the framework of other gravitational theories including the case that the covariant derivative of the energy-momentum tensor is not zero.

Rastall gravity is a gravitational theory established by modifying the vanishing of the covariant derivative of the energy-momentum tensor to be nonvanishing. In this gravitational theory, the  covariant  derivative of the energy-momentum tensor is directly proportional~\cite{P3} to the derivative of the Ricci scalar, i.e. $T^{\mu \nu }_{~\ ~\ ;\mu }\propto R^{,\nu }$. In this respect, Rastall gravity can be seen~\cite{P4} as a phenomenological implementation of some quantum effects in the curved spacetime background. In recent years, Rastall gravity has been applied to cosmology.  It was found~\cite{P5,P6} that Rastall gravity theory is consistent with various observational data in the cosmological context and it gives some new and interesting results at the cosmological level. For instance, the evolution of small dark matter fluctuations is the same as that in the $\mathrm{\Lambda CDM}$ model. But the dark energy is clustered in Rastall theory. This characteristic leads~\cite{P7} to inhomogeneities in the evolution of dark matter in a nonlinear region, which is different from the standard CDM model. 
In addition, it was debating whether Rastall gravity and Einstein gravity are equivalent or not. The inequivalence between the two theories of gravity was pointed out~\cite{PX} several decades ago. 
Nontheless, the equivalence was claimed~\cite{PY} recently, but soon afterwards the inequivalence survived~\cite{PZ} from the point of view that Rastall gravity is a more open theory of gravity than Einstein gravity.

According to string theory, the basic elements of the nature are not point particles but extended one-dimensional objects. Therefore, it is essential to understand what are the gravitational effects induced by a collection of strings. The first study of a cloud of strings as the source of the gravitational field gave~\cite{P8} the exact integral expression of the general solution of the Schwarzschild black hole surrounded by a spherically symmetric string clouds under Einstein gravity, and the following work focused on~\cite{P9,P10} the construction of black hole solutions. So it is natural to explore the gravitational effects of strings as basic objects in a gravitational theory that goes beyond Einstein gravity. For example, the 5-dimensional and $n$-dimensional black holes surrounded by a cloud of strings in Lovelock gravity have been analyzed~\cite{P11,P12}.

The investigation of black hole area and entropy quanta has an important physical meaning because it can provide~\cite{P13} a window to find an effective way to quantize gravity. Bekenstein was the first to suggest~\cite{P14} that the area of black holes should be quantized. That is, if the black hole horizon area was dealt with as an adiabatic invariant, the area spectrum of black holes was proved to be equidistant and quantized in units $\Delta A=8\pi\hbar$. There are a lot of papers about the area spectrum and entropy spectrum of black holes, see, for instance, Refs.~\cite{P13,P15,P16,P17,P18,P19,P20,P21}.

In this paper, our aim is to analyze the quasinormal modes of a Schwarzschild black hole surrounded by a cloud of strings in Rastall gravity. Therefore, we first work out the exact solution of such a static spherically symmetric black hole. Then, we make a simple analysis of the characteristics of the black hole model. According to the method of Regge and Wheeler~\cite{P26}, we derive the quasinormal modes of the odd parity gravitational perturbation for this black hole model and calculate the corresponding quasinormal mode frequencies by using the high order WKB-Pad\'{e} approximation~\cite{ P22,P23}. In the following context, we focus on the influence of a cloud of strings on the real and imaginary parts of the quasinormal mode frequencies. We also utilize  the unstable null geodesics of black holes~\cite{PVC} to compute the quasinormal mode frequencies at the eikonal limit for this black hole model. In addition, we adopt the method of the adiabatic invariant integral~\cite{P13} and that of the periodic property of outgoing waves~\cite{P21} to derive the area spectrum and entropy spectrum of the black hole model, respectively, so as to give the influence of Rastall gravity and of strings on the area spectrum and entropy spectrum.

The paper is organized as follows. In Sect. 2, we first find out the exact solution of a static spherically symmetric black hole surrounded by a cloud of strings in Rastall gravity. Then, we analyze in Sect. 3 the characteristics of the black hole model for different values of the two parameters related to Rastall gravity and strings, where the case of the Einstein theory is added as a contrast. In Sect. 4, we calculate the quasinormal mode frequencies for this black hole model in terms of the two different strategies mentioned above and give the consistent conclusions. In Sect. 5, we use the method of the adiabatic invariant integral to compute the area spectrum and entropy spectrum of the black hole model, and then we utilize the method of the periodic property of outgoing waves to calculate the same quantities and obtain the same results  in Sec. 6. Finally, we make a simple summary in Sect. 7. Through out this paper, we adopt  the units $c=G=k_{B}=1$ and the sign convention $(+,-,-,-)$.

\section{Schwarzschild black hole surrounded by a cloud of strings in Rastall gravity}
According to  Rastall gravity, the field equations take~\cite{P3} the following forms,
\begin{equation}
\label{1}
G_{\mu \nu}+\beta g_{\mu \nu}R=\kappa T_{\mu \nu},
\end{equation}
\begin{equation}
\label{2}
{T^{\mu \nu }}_{;\mu}=\lambda R^{,\nu},
\end{equation}
where $\beta \equiv \kappa \lambda$, $\lambda$ is the Rastall parameter and $\kappa$  is the Rastall gravitational coupling constant. From Eq.~(\ref{1}) and Eq.~(\ref{2}) one can get 
\begin{equation}
\label{3}
R=\frac{\kappa }{4\beta-1}T,
\end{equation}
\begin{equation}
\label{4}
{T^{\mu \nu }}_{;\mu }=\frac{\beta }{4\beta-1}T^{,\nu},
\end{equation}
where $R$ is the Ricci scalar and $T={T^{\mu}}_{\mu}$ is the trace of the energy-momentum tensor. The  Newtonian  limit~\cite{P24} leads to $\kappa =\frac{4\beta -1}{6\beta -1}8\pi $. Obviously,  when  $\lambda =0$, i.e. $\beta=0$, $\kappa$ is equal to $8\pi $, which gives rise to the recovery of Einstein gravity and the conservation of the  energy-momentum tensor. It should be noted that $\beta =\frac{1}{6}$ and $\beta =\frac{1}{4}$ are not allowed~\cite{P24}.

In order to derive the solution of a Schwarzschild black hole surrounded by a cloud of strings in Rastall theory, we consider the following static spherically symmetric metric,
\begin{equation}
\label{5}
ds^{2}=f(r)dt^{2}-f^{-1}(r)dr^{2}-r^{2}(d\theta^{2}+\sin^{2}\theta d\phi ^{2}),
\end{equation}
and then compute the non-vanishing components of the Rastall tensor defined by $H_{\mu \nu }\equiv G_{\mu \nu }+\beta g_{\mu \nu }R$ as follows:
\begin{equation}
\label{6}
{H^{0}}_{0 }={G^{0}}_{0 }+\beta R=-\frac{rf'(r)+f(r)-1}{r^{2}}+\beta R,
\end{equation}
\begin{equation}
\label{7}
{H^{1}}_{1}={G^{1}}_{1}+\beta R=-\frac{rf'(r)+f(r)-1}{r^{2}}+\beta R,
\end{equation}
\begin{equation}
\label{8}
{H^{2}}_{2}={G^{2}}_{2}+\beta R=-\frac{rf''(r)+2f'(r)}{2r}+\beta R,
\end{equation}
\begin{equation}
\label{9}
{H^{3}}_{3}={G^{3}}_{3}+\beta R=-\frac{rf''(r)+2f'(r)}{2r}+\beta R,
\end{equation}
where $R=[(r^{2}f''(r)+4rf'(r)+2f(r)-2)]/r^{2}$, and $f'(r)$ and $f''(r)$ stand for the first and second derivatives of $f(r)$ with respect to $r$, respectively.

The first solution of a static spherically symmetric black hole with a cloud of strings in Einstein gravity was given by Letelier~\cite{P8} in which the action of a string in the evolution of spacetime reads
\begin{equation}
\label{16}
S=\int _{\Sigma }m\sqrt{-\gamma }d\xi ^{0}d\xi ^{1},
\end{equation}
where $m$ is a dimensionless constant that is related to the tension of the string. And $\gamma$ is the determinant of the induced metric,
\begin{equation}
\label{17}
\gamma_{ab}=g_{\mu \nu} \frac{\partial x^{\mu}}{\partial\xi^{a}}\frac{\partial x^{\nu}}{\partial \xi^{b}},
\end{equation}
where $x^{\mu }=x^{\mu }(\xi^{a})$ describes a two-dimensional string world sheet $\Sigma$, 
and $\xi ^{0}$ and $\xi ^{1}$ are timelike and spacelike parameters, respectively.
The string bivector which is associated to the string world sheet is defined by
\begin{equation}
\label{18}
\Sigma ^{\mu \nu }=\epsilon ^{ab}\frac{\partial x^{\mu }}{\partial\xi^{a}  }\frac{\partial x^{\nu }}{\partial \xi^{b}},
\end{equation}
where $\epsilon ^{ab}$ is the two-dimensional Levi-Civita symbol with $\epsilon ^{01}=-\epsilon ^{10}=1$.

Now we consider a cloud of strings with world sheets. The energy-momentum tensor of a cloud of strings characterized by a proper density $\rho $ reads 
\begin{equation}
\label{19}
T^{\mu \nu }=\frac{\rho \Sigma ^{\mu \beta }{\Sigma_{\beta }}^{\nu }}{\sqrt{-\gamma }},
\end{equation}
where $\gamma =\frac{1}{2}\Sigma ^{\mu \nu }\Sigma _{\mu \nu }$. 
Because the string cloud is spherically symmetric, this restricts the density $\rho $ and the string bivector $\Sigma ^{\mu \nu }$ to be only a function of $r$. In this situation, the only non-vanishing components
of the string bivector are $\Sigma ^{01}$ and $\Sigma ^{10}$, which are  linked by $\Sigma ^{01}=-\Sigma ^{10}$. Thus the only non-vanishing components of the energy-momentum tensor of a cloud of strings take the forms,
\begin{equation}
\label{20}
{T^{0}}_{0}={T^{1}}_{1}=-\rho \Sigma ^{01}.
\end{equation}
That is, the energy-momentum tensor of a cloud of strings with the spherical symmetry can be written as 
\begin{equation}
\label{21}
{T^{ \mu}}_{\nu }=\begin{pmatrix}
\rho _{s}(r) &  &  &  &\\ 
 &  \rho _{s}(r) &  & &\\ 
 &  &  0 & &\\ 
 &  &  & 0 &\\
\end{pmatrix}.
\end{equation}
After substituting Eq.~(\ref{21}) into Eq.~(\ref{4}), we can get the following equation,
\begin{equation}
\label{22}
\frac{\mathrm{d} \rho _{s}}{\mathrm{d} r}+\frac{2\rho _{s}}{r}=\frac{2\beta }{4\beta -1}\frac{\mathrm{d} \rho _{s}}{\mathrm{d} r},
\end{equation}
from which we deduce the solution,
\begin{equation}
\label{23}
\rho _{s}(r)=\frac{b}{r^{\frac{2(4\beta -1)}{2\beta -1}}},
\end{equation}
where $b$ is an integral constant linked to the cloud of strings.

Considering the line element given by Eq.~(\ref{5}), we rewrite the Rastall field equations as follows:
\begin{equation}
\label{28}
 -\frac{rf'(r)+f(r)-1}{r^{2}}+\beta R=\kappa\rho _{s},
\end{equation}
\begin{equation}
\label{29}
-\frac{rf''(r)+2f'(r)}{2r}+\beta R=0,
\end{equation}
and obtain the general solution of the metric function,
\begin{equation}
\label{301}
f(r)=1-\frac{C_{1}}{r}+r^{\frac{1-2\beta }{\beta }}C_{2}+\frac{4\kappa b(\beta -\frac{1}{2})^{2}}{(8\beta ^{2}+2\beta -1)r^{\frac{4\beta}{2\beta -1}}},
\end{equation}
where $C_{1}$ and $C_{2}$ are two constants to be determined.

When $\beta =b=0$, Eq.~(\ref{301}) should go back to the Schwarzschild vacuum solution in Einstein gravity. Moreover, Eq.~(\ref{301}) should not be divergent at $\beta =0$. The two constraints require $C_{1}=2M$ and $C_{2}=0$. Therefore, we reach the final form of the function,
\begin{equation}
\label{30}
f(r)=1-\frac{2M}{r}+\frac{4a(\beta -\frac{1}{2})^{2}}{(8\beta ^{2}+2\beta -1)r^{\frac{4\beta }{2\beta -1}}},
\end{equation}
where the parameter $a$ is defined as $a\equiv \kappa b$. It is easy to check that Eq.~(\ref{30}) turns back~\cite{P8} to the case of Einstein gravity when $\beta=0$. In addition, when $\beta <\frac{1}{6}$ or $\beta >\frac{1}{4}$, $\kappa =\frac{4\beta -1}{6\beta -1}8\pi$ is positive, so that $a$ is proportional to $b$ and represents the effect of strings.

\section{Characteristics of the black hole model}

According to Eq.~(\ref{30}), we can compute the radius of the event horizon from the equation  $f(r_{\rm H})=0$, which gives the relationship between the black hole mass and the radius,
\begin{equation}
\label{32}
M=\frac{1}{2} r_{\rm H} \left(1+\frac{4 a \left(\beta -\frac{1}{2}\right)^2 r_{\rm H}^{-\frac{4 \beta }{2 \beta -1}}}{8 \beta ^2+2 \beta -1}\right).
\end{equation}
Using $\kappa_{r_{\rm H}}= \left.\frac{1}{2}\frac{\mathrm{d} f(r)}{\mathrm{d} r}\right|_{r=r_{\rm H}}$, we obtain the surface gravity of the Schwarzschild black hole surrounded by a cloud of strings in Rastall gravity,
\begin{equation}
\label{33}
\kappa_{r_{\rm H}}=\frac{1}{2 r_{\rm H}}+\frac{a (1-2 \beta ) r_{\rm H}^{\frac{4 \beta }{1-2 \beta }}}{2 (4 \beta -1) r_{\rm H}}.
\end{equation}
Then we can calculate the corresponding  Hawking temperature through the surface gravity as follows,
\begin{equation}
\label{34}
T_{\rm {BH}}=\frac{\hbar\kappa_{r_{\rm H}}}{2\pi }=\frac{\hbar}{4\pi r_{\rm H} }\left(1+\frac{a (1-2 \beta ) r_{\rm H}^{\frac{4 \beta }{1-2 \beta }}}{4 \beta -1}\right).
\end{equation}

In the following, we analyze the characteristics of the black hole model for different values of the two parameters $\beta$ and $a$, where $\beta$ represents the effect of Rastall gravity and $a$ that of strings.

In Fig. 1.1 and Fig. 1.2, we draw the graphs of function $f(r)$ with respect to $r$ for different values of $a$ when $\beta >0$ and $\beta <0$, respectively. It should be noted that the curves with $\beta =0$ in Fig. 1.1 and Fig. 1.2 represent the case in Einstein gravity, which is given as a contrast. From Fig. 1.1, we can see that when $\beta >0$, $f(r)$ in Rastall gravity is significantly lower than that in Einstein gravity at the same value of $a$. For example, when the parameter $a=0.1$, the red curve representing Rastall gravity is significantly lower than the black curve representing Einstein gravity in Fig. 1.1. Moreover, as the value of $a$ increases, $f(r)$ in Rastall gravity keeps decreasing. When $a$ is small, there are generally two  horizons as shown by the red curve in Fig. 1.1, where the first is the event horizon, and the second is what we call the string horizon which is similar to the quintessence horizon of the  spherically symmetric black hole  surrounded by quintessence in Ref.~\cite{PMa}. However, the two horizons will shrink to one or even no horizons appear when $a$ is large, see, for instance, the blue curve for the former case and the green curve for the latter in Fig. 1.1. From Fig. 1.2, we can see that when $\beta <0$, $f(r)$ in Rastall gravity is significantly higher than that in Einstein gravity at the same value of $a$. 
For example, when we compare the two curves in red and black ($a=0.1$) and the two curves in green and purple ($a=0.9$), we notice that the red curve representing Rastall gravity is significantly higher than the black curve representing Einstein gravity in Fig. 1.2, and the same situation exists for the green and purple curves.  
Moreover, as the value of $a$ increases, $f(r)$ in Rastall gravity also keeps decreasing, which is same as the case of $\beta >0$, see the red, blue and green curves of Fig. 1.2. 
Nonetheless, the black hole still maintains one event horizon in Rastall gravity when $a$ is large, see, for instance, the green curve in Fig. 1.2, which is quite different from the case of $\beta >0$.

\begin{figure}
\centering
\begin{minipage}[t]{0.8\linewidth}
\includegraphics[width=140mm]{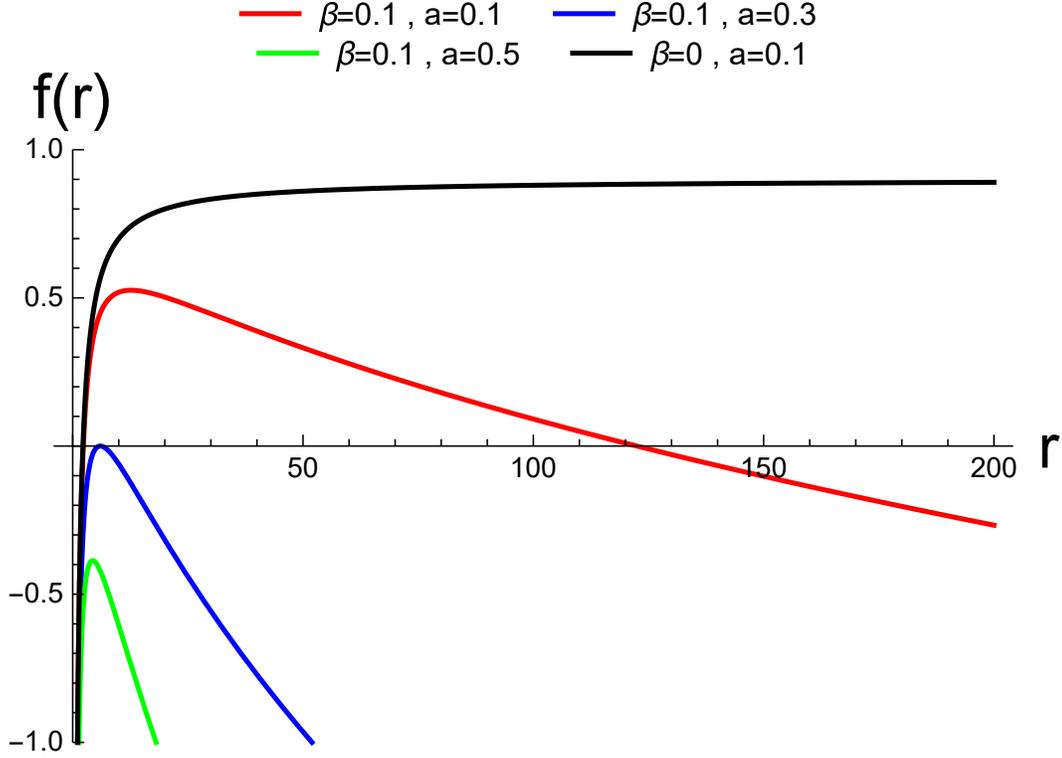}
\caption*{Fig. 1.1 Graph of function $f(r)$ with respect to $r$  for different values of $a$. Here we choose M = 1  and $\beta =0.1$, and add $\beta =0$ as a contrast.}
\label{fig11} 
\end{minipage}
\end{figure}

\begin{figure}
\centering
\begin{minipage}[t]{0.8\linewidth}
\includegraphics[width=140mm]{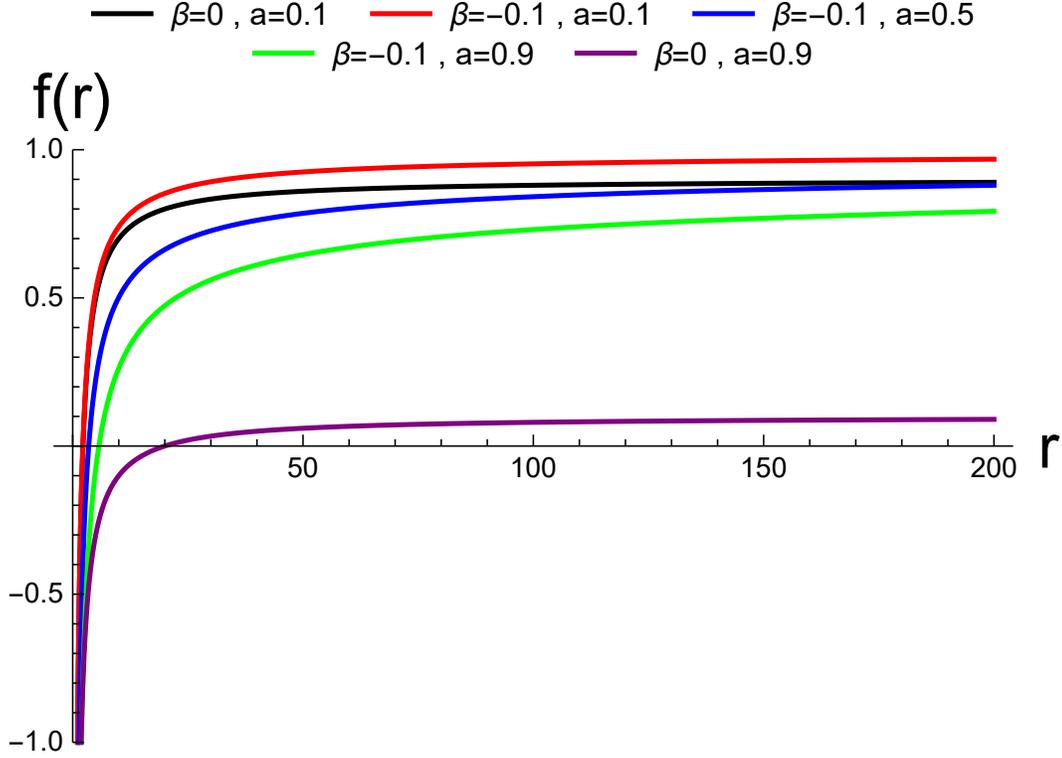}
\caption*{Fig. 1.2 Graph of function $f(r)$ with respect to $r$  for different values of $a$. Here we choose M = 1 and $\beta =-0.1$, and add $\beta =0$ as a contrast.}
\label{fig12} 
\end{minipage}
\end{figure}

In Fig. 2.1 and Fig. 2.3, we draw the graphs of the Hawking temperature $T_{\rm {BH}}$ with respect to the event horizon radius $r_{\rm H}$ for different values of $a$ when $\beta >0$ and $\beta <0$, respectively. In Fig. 2.2 and Fig. 2.4, we draw the graphs of the Hawking temperature $T_{\rm {BH}}$ with respect to the event horizon radius $r_{\rm H}$ for different values of $a$ and $\beta$ in Rastall gravity and in Einstein gravity when $\beta >0$ and $\beta <0$, respectively. From Fig. 2.1 and Fig. 2.2, we can see that $T_{\rm {BH}}$ in Rastall gravity keeps decreasing as the value of $a$ increases, see the curves in red, blue and green.  Moreover, when $\beta >0$, at the same value of $a$, $T_{\rm {BH}}$ in Rastall gravity is higher than that in Einstein gravity at a small $r_{\rm H}$ and then lower than that in Einstein gravity at a large $r_{\rm H}$, see the three pairs of curves in red and black, in blue and orange, and in green and purple, respectively, in Fig. 2.2. From Fig. 2.3 and Fig. 2.4, we can see that $T_{\rm {BH}}$ in Rastall gravity also keeps decreasing as the value of $a$ increases.  Moreover, when $\beta <0$, at the same value of $a$, $T_{\rm {BH}}$ in Rastall gravity is lower than that in Einstein gravity at a small $r_{\rm H}$ and then higher than that in Einstein gravity at a large $r_{\rm H}$, see the three pairs of curves in red and black, in blue and orange, and in green and purple, respectively, in Fig. 2.4.

In addition, we find that when $\beta >0$, $T_{\rm {BH}}$ decreases monotonically with respect to the increase of $r_{\rm H}$, see Fig. 2.1 and Fig. 2.2, which means that the black hole temperature will increase continuously during the whole stage of evaporation, especially the Hawking temperature $T_{\rm {BH}}$ will diverge when $r_{\rm H}$ tends to zero. However, when $\beta <0$ and the parameter $a$ takes a suitable value, say, $a=0.5$ which corresponds to the blue curves in Fig. 2.3 and Fig. 2.4, $T_{\rm {BH}}$ increases at first and then decreases, and finally tends to zero with respect to the increase of $r_{\rm H}$, which means that at the initial state of evaporation, the black hole temperature increases with respect to the decrease of $r_{\rm H}$, and after the black hole temperature increases to a finite maximum, it quickly drops to zero at $r_{\rm H}=r_{0}$ (the event horizon radius of the extreme black hole), leaving a frozen black hole. The former case ($\beta >0$) is similar to the situation of a classical Schwarzschild black hole in Einstein gravity, while the latter ($\beta <0$) is similar to the situation of a noncommutative Schwarzschild black hole~\cite{PAE} in Einstein gravity.

\begin{figure}
\centering
\begin{minipage}[t]{0.80\linewidth}
\centering
\includegraphics[width=140mm]{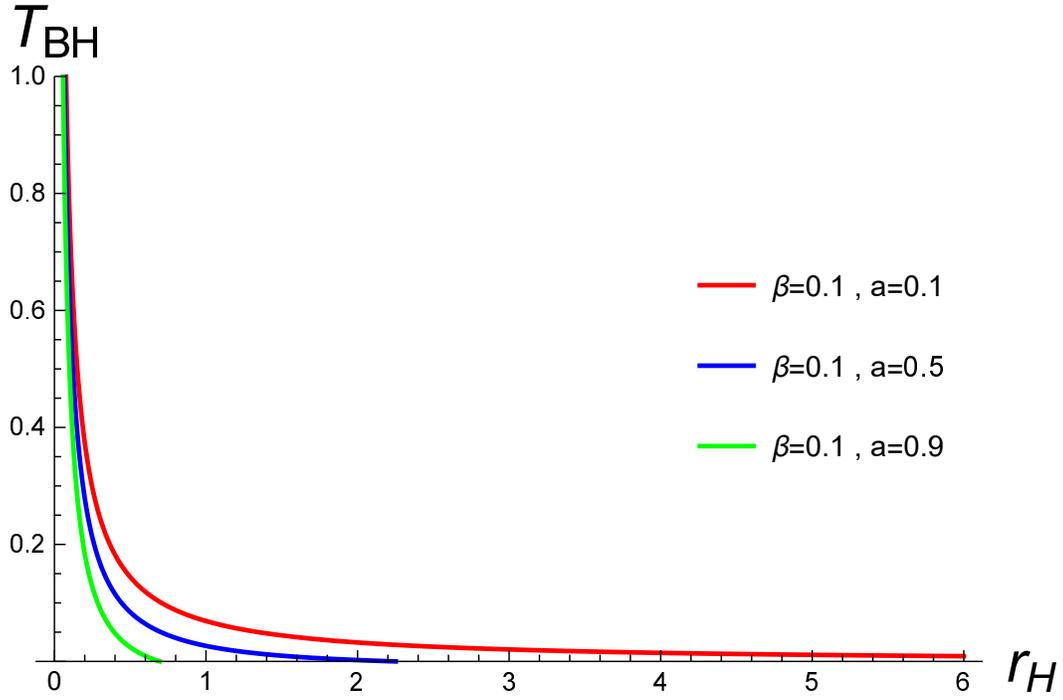}
\caption*{Fig. 2.1 Graph of the Hawking temperature $T_{\rm {BH}}$ with respect to the event horizon radius $r_{\rm H}$ for different values of $a$. Here we  choose $M = \hbar=1$ and $\beta =0.1$.}
\label{fig21} 
\end{minipage}
\end{figure}

\begin{figure}
\centering
\begin{minipage}[t]{0.80\linewidth}
\centering
\includegraphics[width=140mm]{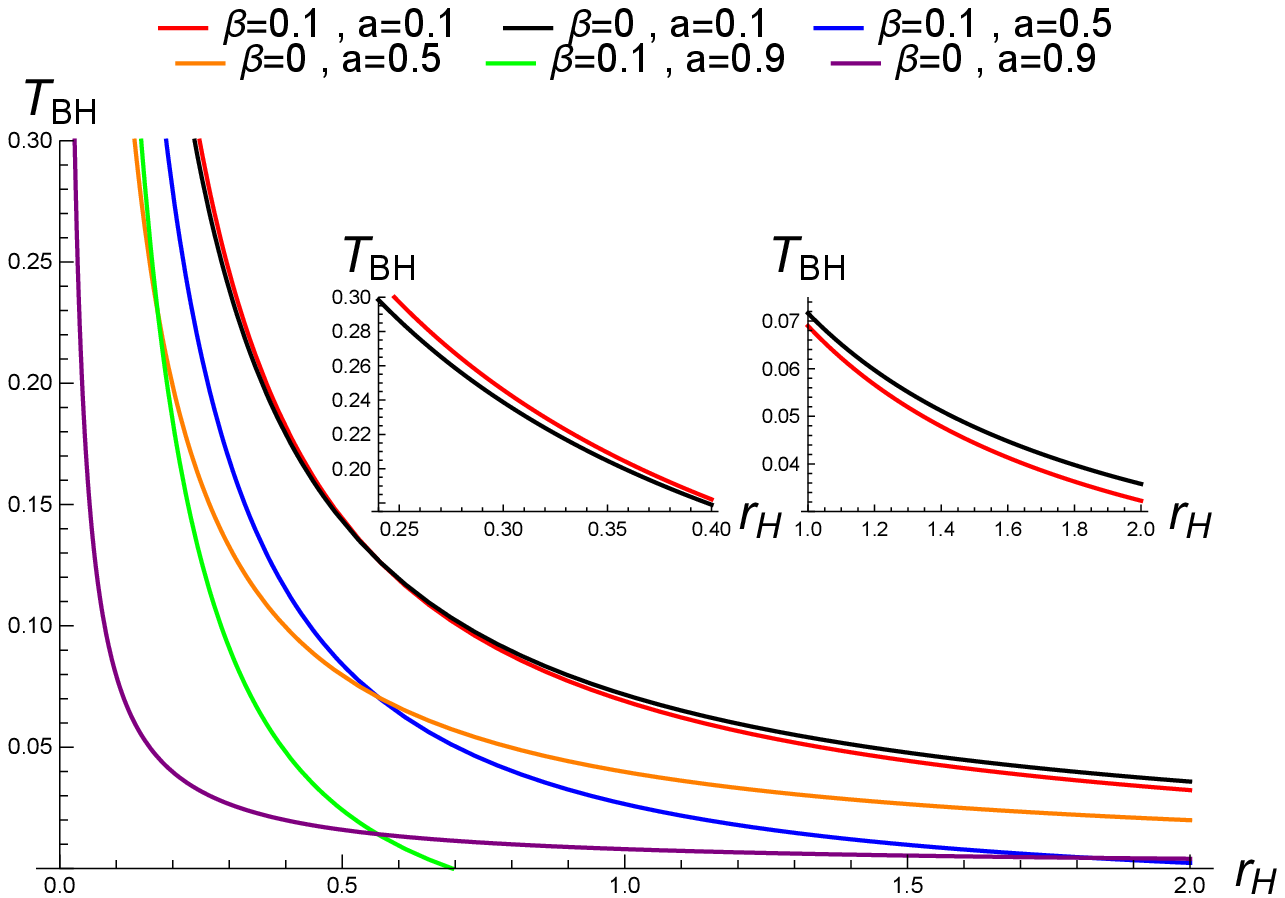}
\caption*{Fig. 2.2 Graph of the Hawking temperature $T_{\rm{BH}}$ with respect to the event horizon radius $r_{\rm H}$ for different values of $a$ and $\beta $. Here we choose $M = \hbar=1$.}
\label{fig22} 
\end{minipage}
\end{figure}

\begin{figure}
\centering
\begin{minipage}[t]{0.80\linewidth}
\centering
\includegraphics[width=140mm]{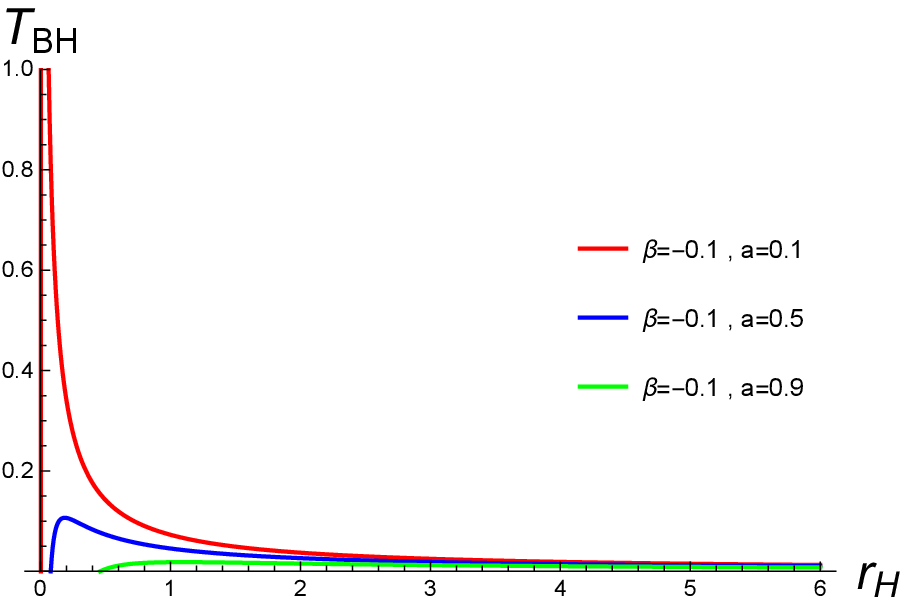}
\caption*{Fig. 2.3 Graph of the Hawking temperature $T_{\rm{BH}}$ with respect  to the event horizon radius $r_{\rm H}$ for different values of $a$. Here we choose  $M = \hbar=1$ and $\beta =-0.1$.}
\label{fig23} 
\end{minipage}
\end{figure}

\begin{figure}
\centering
\begin{minipage}[t]{0.80\linewidth}
\centering
\includegraphics[width=140mm]{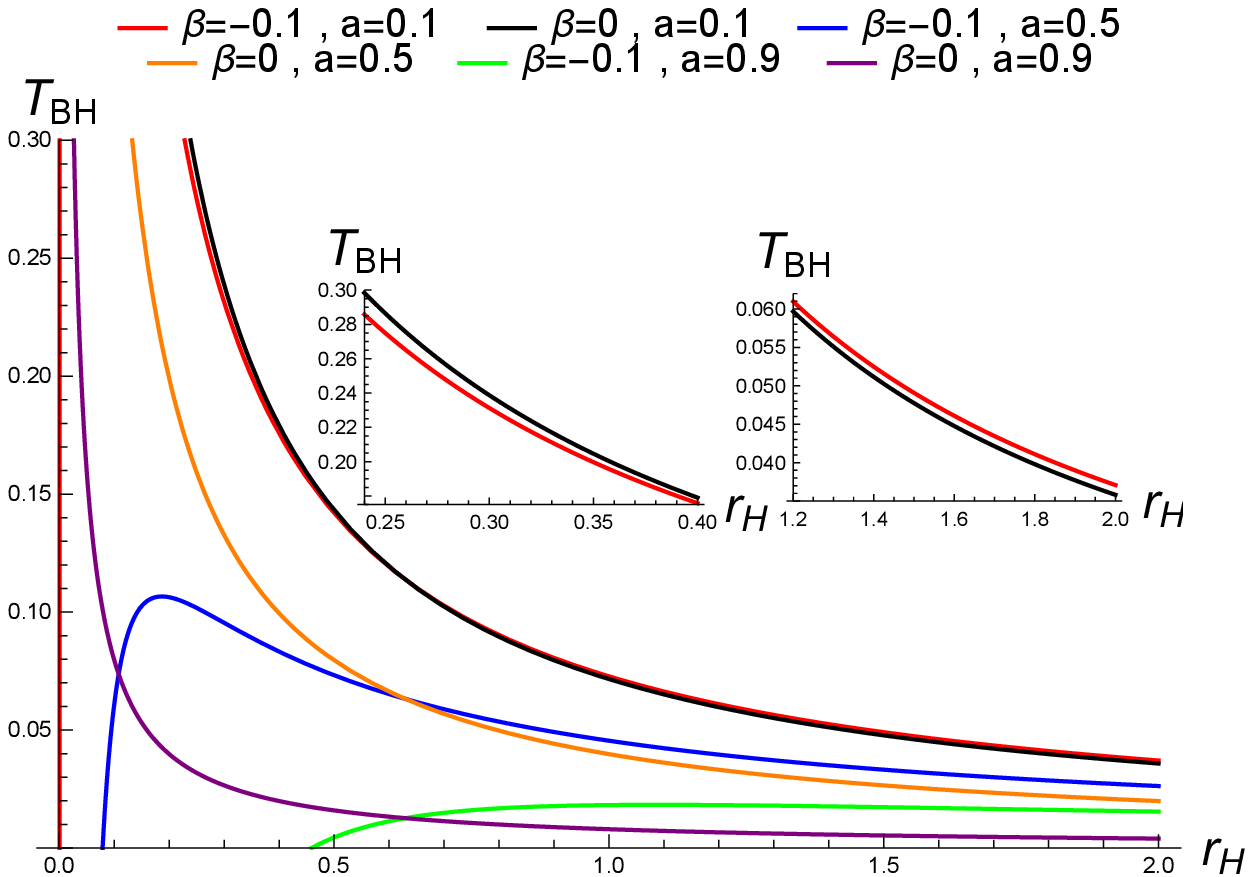}
\caption*{Fig. 2.4 Graph of the Hawking temperature $T_{\rm {BH}}$ with respect to the event horizon radius $r_{\rm H}$ for different values of $a$ and $\beta $. Here we choose $M = \hbar=1$.}
\label{fig24} 
\end{minipage}
\end{figure}

At the end of this section, we try to establish the relationship between the solution of the static spherically symmetric black hole surrounded by a cloud of strings in Rastall gravity and the solution of the static spherically symmetric black hole surrounded by quintessence in Einstein gravity. We find that the two solutions exchange to each other when their parameters are connected to each other in a specific formula. Therefore, we provide a possibility for the string fluid to be a candidate of dark energy if the parameter $\beta$ of Rastall gravity is positive. Such a possibility has been enhanced by a recent work~\cite{PLRWJ} in which the Rastall parameter $\beta$ was fixed to be 0.163 through the analyses of 118 galaxy-galaxy strong gravitational lensing systems.

For a static spherically symmetric solution of Einstein's equations describing a black hole surrounded by the quintessential matter under the condition of additivity and linearity in energy-momentum tensor, the metric takes~\cite{P25} the following form,
\begin{equation}
\label{63}
ds^{2}=\left(1-\frac{2M}{r}-\frac{\tilde c}{r^{3\omega _{q}+1}}\right)dt^{2}-\left(1-\frac{2M}{r}-\frac{\tilde c}{r^{3\omega _{q}+1}}\right)^{-1}dr^{2}-r^{2}(d\theta^{2}+\sin^{2}\theta d\phi ^{2}),
\end{equation}
where  $M$ is the black hole mass, $\omega _{q}$ is the quintessential state parameter, and $\tilde c$ is an integral constant linked to the energy density of quintessence.
We compare Eq.~(\ref{30}) with Eq.~(\ref{63}) and see that the solution of the static spherically symmetric black hole surrounded by a cloud of strings in Rastall gravity can be transformed into the solution of the static spherically symmetric black hole surrounded by quintessence in Einstein gravity. The specific transformation formula is as follows:

\begin{equation}
\label{64}
\beta =\frac{3 \omega _q+1}{6 \omega _q-2},  
\end{equation}

\begin{equation}
\label{65}
a=-\frac{9\tilde c}{2}\left(\omega _q+\omega _q^2\right).
\end{equation}
Now we draw the graphs of $\beta$ and ${a}/{\tilde c}$ as a function of $\omega _q$. Note that the range of parameter $\omega _q$ related to the quintessence field is constrained~\cite{P25} from $-1$ to $-\frac{1}{3}$.

In Fig. 3.1 and Fig. 3.2, we draw the graphs of the parameter $\beta $ and the parameter ${a}/{\tilde c}$ with respect to the quintessential state parameter $\omega _{q}$, respectively. From Fig. 3.1, we can see that $\beta $ decreases monotonically with the increase of $\omega _{q}$. From Fig. 3.2, we can see that ${a}/{\tilde c}$ increases at first and then decreases with the increase of $\omega _{q}$, where it reaches the maximum, ${a}/{\tilde c}=1.125$, when $\omega _{q}=-0.5$.

\begin{figure}
\centering
\begin{minipage}[t]{0.6\linewidth}
\includegraphics[width=105mm]{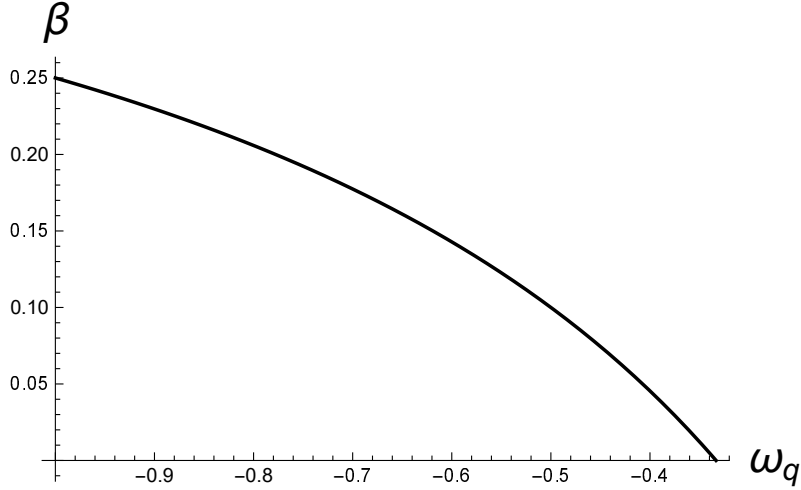}
\caption*{Fig. 3.1 Graph of the parameter $\beta $ with respect to  the  quintessential state parameter $\omega _{q}$.}
\label{fig31} 
\end{minipage}
\end{figure}

\begin{figure}
\centering
\begin{minipage}[t]{0.60\linewidth}
\includegraphics[width=105mm]{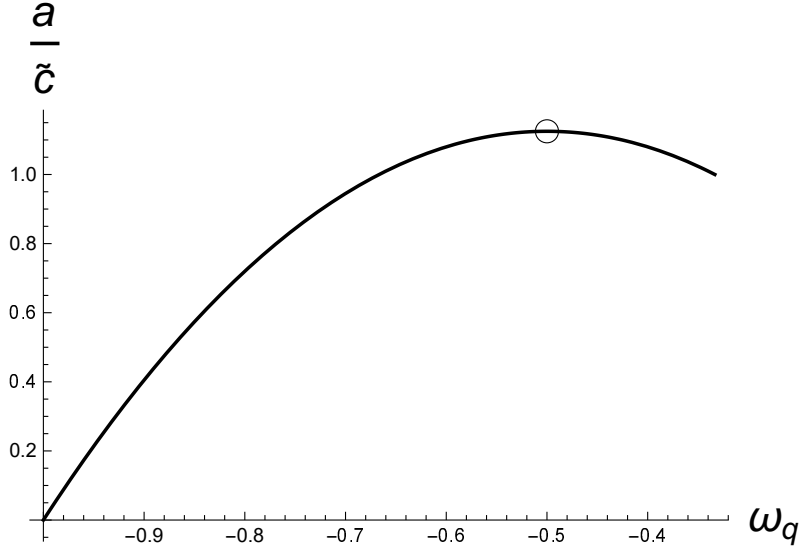}
\caption*{Fig. 3.2 Graph of the parameter ${a}/{\tilde c}$ with respect to the quintessential state parameter $\omega _{q}$. The circle in the figure represents the highest point ($-0.5$, $1.125$).}
\label{fig32} 
\end{minipage}
\end{figure}

\section{Quasinormal mode frequency of gravitational perturbation}
The gravitational perturbation of black holes was first studied by Regge and Wheeler~\cite{P26} for the odd parity type of the spherical harmonics, and then extended to the even parity type by Zerilli~\cite{P27}.
At present, the odd parity gravitational perturbation is utilized in a large number of papers, especially for the black holes surrounded by quintessence, in terms of the method of Regge and Wheeler dealing with the vacuum case, see, for instance, Refs.~\cite{YZYXG,MSBBTC,MSTBBK,PQ}. Therefore, we follow the same method to calculate the quasinormal mode frequencies in our model, i.e. the Schwarzschild black hole surrounded by a cloud of strings in Rastall gravity. In addition, in order to increase the reliability of our calculations, we use alternatively the unstable null geodesics to compute once more the quasinormal mode frequencies at the eikonal limit for this black hole model. Since the scalar, electromagnetic and gravitational perturbations in the background of the static spherically symmetric spacetime have the same behavior under the eikonal limit, the corresponding results can be regarded as a reference standard for our results obtained by the method of Regge and Wheeler together with the high order WKB-Pad\'e approximation. We find that the conclusions made by the two strategies (one is the method of Regge and Wheeler and the other the unstable null geodesics at the eikonal limit) are consistent.  

In the gravitational perturbation, $g_{\mu\nu}$ usually represents the background metric and $h_{\mu\nu}$ the perturbation. As $h_{\mu\nu}$ is very small when it compares to $g_{\mu\nu}$, the canonical form for the odd parity perturbation reads
\begin{equation}
\label{56}
h_{\mu\nu}=\begin{vmatrix}
0 & 0 &  0 &  h_{0}(r) &\\ 
0 & 0 & 0 &  h_{1}(r) &\\ 
0 & 0 &  0 &  0 & \\ 
 h_{0}(r) &  h_{1}(r) & 0 & 0 &
\end{vmatrix}\mathrm{exp}(-i\omega t)\mathrm{sin\theta }\frac{\partial }{\partial \theta }P_{l}(\mathrm{cos\theta}),
\end{equation}
where $l$ is the angular quantum number, $\omega $ the complex quasinormal mode frequency, $P_{l}(\mathrm{cos\theta})$ the Legendre function, and $h_{0}(r)$ and $h_{1}(r)$ two independent components of $h_{\mu\nu}$. We now derive the desired 
Schr\"odinger-like equation, i.e. the master equation.

One can compute $R_{\mu \nu }$ from $g_{\mu \nu }$ and similarly calculate $R_{\mu \nu }+\delta R_{\mu \nu }$ from $g_{\mu \nu }+h_{\mu \nu } $, and thus deduce~\cite{M1} $\delta R_{\mu \nu }$ as follows:

\begin{equation}
\label{57}
\delta R_{\mu \nu }=-\delta \Gamma _{\mu \nu ;\beta }^{\beta }+\delta \Gamma _{\mu \beta ;\nu }^{\beta },
\end{equation}
where 
\begin{equation}
\label{58}
\delta \Gamma _{\beta \gamma }^{\alpha }=\frac{1}{2}g^{\alpha \nu }(h_{\beta \nu ;\gamma }+h_{\gamma \nu ;\beta }-h_{ \beta \gamma  ;\nu }).
\end{equation}
By substituting Eq.~(\ref{56}) into Eq.~(\ref{57}), we obtain two independent perturbation field equations for $h_{0}(r)$ and $h_{1}(r)$,
\begin{equation}
\label{59}
\frac{i\omega h_{0}(r)}{f(r)}+\frac{\mathrm{d}(f(r)h_{1}(r)) }{\mathrm{d} r}=0,   
\end{equation}
\begin{equation}
\label{60}
\frac{i\omega }{f(r)}\left(\frac{\mathrm{d} h_{0}(r)}{\mathrm{d} r}+i\omega h_{1}(r)-\frac{2h_{0}(r)}{r}\right)+(l-1)(l+2)\frac{h_{1}(r)}{r^{2}}=0,  
\end{equation}
where the former corresponds to $\delta R_{23}=0$ and the latter to $\delta R_{13}=0$, respectively, and $f(r)$ has been given in Eq.~(\ref{30}). Eliminating $h_{0}(r)$ by $h_{1}(r)$ in the above two equations, and then defining $\Psi(r) =\frac{f(r)h_{1}(r)}{r}$ and $\frac{\mathrm{d} r_{*}}{\mathrm{d} r}=\frac{1}{f(r)}$, 
we finally reach the master equation,
\begin{equation}
\label{61}
\frac{\mathrm{d}^{2} \Psi(r) }{\mathrm{d} r_{*}^{2}}+[\omega ^{2}-V(r)]\Psi(r) =0,
\end{equation}
where the effective potential $V(r)$ reads
\begin{equation}
\label{62}
V(r)=\left(1-\frac{2 M}{r}+\frac{4 a \left(\beta -\frac{1}{2}\right)^2}{\left(8 \beta ^2+2 \beta -1\right) r^{\frac{4 \beta }{2 \beta -1}}}\right)\left(\frac{l (l+1)}{r^2}-\frac{6 M}{r^3}-\frac{2 a r^{\frac{4 \beta }{1-2 \beta }-2}}{2 \beta +1}+\frac{4 a \beta  r^{\frac{4 \beta }{1-2 \beta }-2}}{2 \beta +1}\right).
\end{equation}

In order to let the effective potential satisfy $V(r)\rightarrow 0$ at $r\rightarrow \infty $, the condition $\beta <\frac{1}{6}$ is needed, see also Ref.~\cite{P24}. In addition, we can easily see that both $f(r)$ and $V(r)$ are divergent when $\beta =-0.5$.
Therefore, we set the range of $\beta$ be $-0.5<\beta <\frac{1}{6}$, which conforms to the entropy positivity condition~\cite{P24} and ensures that the parameter $a$ is proportional to the parameter $b$. As to the range of $a$, it depends on the positivity or negativity of $\beta$. For $\beta>0$, the black hole will have no event horizons when $a$ is large. Thus, we set the range of $a$ be $0\leqslant a\leqslant 0.30$ for $\beta=0.1$. For $\beta<0$, the barrier height of the effective potential will disappear when $a$ goes to one, so we take the range of values to be $0\leqslant a< 1$. 

In the following, we use the high order WKB-Pad\'e approximation to calculate the quasinormal mode frequencies of the odd parity gravitational perturbation, and alternatively apply the unstable null geodesics to compute the quasinormal mode frequencies at the eikonal limit for this black hole model when $\beta>0$ and $\beta<0$, respectively.

\subsection{Quasinormal mode frequencies calculated by  the high order WKB-Pad\'e approximation  } 

As for the high order WKB-Pad\'e approximation, it was first proposed~\cite{P22} by  Matyjasek and  Opala and then developed by  Konoplya et al.~\cite{P23} through a special averaging treatment. In this treatment, the quantity $\Delta _{k}=\frac{|\omega _{k+1}-\omega _{k-1}|}{2}$ is defined for the error estimation and it is positively correlated with the relative error $E _{k}=\left | \frac{\omega _{k}-\omega}{\omega } \right |\times 100\%$ of the quasinormal mode frequencies, where $k$ stands for the order number, $\omega_{k}$ the $k$-th order quasinormal mode frequency calculated by the WKB approximation or the WKB-Pad\'e approximation, and $\omega$ the accurate value of the quasinormal mode frequency. In our calculations, we at first give the quasinormal mode frequencies from the 1st to 13th orders, and then work out the error estimations of the 13 orders' frequencies, and finally pick out such an order's frequency that has the smallest relative error. The results are shown in Table 1 and Table 2 for $\beta>0$, and in Table 3 and Table 4 for $\beta<0$. Here we have two notes: 

(1) We find that when $\beta$ is taken to be a fraction, the result calculated by the high order WKB-Pad\'e approximation is better than that when $\beta$ is a decimal.\footnote{The main reason lies in the term $r^{\frac{4\beta }{2\beta -1}}$ in the metric function $f(r)$. When $\beta$ takes a decimal, it usually makes the power of $r^{\frac{4\beta }{2\beta -1}}$ a decimal, too. For example, when $\beta=0.0400$, the power equals $-0.1739$, which makes it impossible to solve the horizon radius analytically.
In addition, the high order WKB-Pad\'e approximation will inevitably cause a  deviation when $f(r)$ containing a term like $r^{-0.1739}$ is substituted numerically into iterative formulas. Fortunately, we find that when $\beta$ takes some specific fractions, for example, $\beta=\frac{1}{22}$, $r^{\frac{4\beta }{2\beta -1}}$ becomes $r^{-\frac{1}{5}}$, which makes $f(r)$ simpler and the result calculated by the high order WKB-Pad\'e approximation better. We note that one may expect to get fine approximations of quasinormal mode frequencies and smooth curves for any values of $\beta$ if the precision can be controlled properly in the numerical calculations.} 
Since the range of $\beta$ is set to be $-0.5<\beta <\frac{1}{6}$, the term $r^{\frac{4\beta }{2\beta -1}}$ in Eq.~(\ref{30}) will change from $r^{0}$ to $r^{-1}$ when $\beta$ changes from $0$ to $\frac{1}{6}$, and it will change from $r^{0}$ to $r^{1}$ when $\beta$ changes from $0$ to $-0.5$. 
Therefore, when $\beta$ is taken to be a suitable series of fractions from $0$ to $\frac{1}{6}$, such as $0$, $\frac{1}{82}$, $\frac{1}{42}$, $\frac{3}{86}$, $\cdots$, $\frac{19}{118}$,
see Table 1, the power of $r^{\frac{4\beta }{2\beta -1}}$ is also a series of fractions rather than decimals: ${0}$, $-{\frac{1}{20}}$, $-{\frac{2}{20}}$, $-{\frac{3}{20}}$,  $\cdots$,  $-{\frac{19}{20}}$.
Similarly, when $\beta$ is taken to be a suitable series of fractions from $0$ to$-0.5$, such as $0$, $-\frac{1}{18}$, $-\frac1 8$, $-\frac {3}{14}$, $-\frac 1 3$, and $-\frac {19}{42}$, see Table 3 and Table 4, the power of $r^{\frac{4\beta }{2\beta -1}}$ is also a series of fractions rather than decimals: ${0}$, ${\frac{1}{5}}$, ${\frac{2}{5}}$, ${\frac{3}{5}}$, ${\frac{4}{5}}$, and ${\frac{19}{20}}$.

(2) The order number with the smallest relative error is not less than 4.

\bibliographystyle{unsrt}   
\bibliography{citedpapers}

\begin{table}[htbp]\small
\centering
\caption{The quasinormal mode frequencies of the gravitational perturbation in a Schwarzschild black hole surrounded by a cloud of strings in Rastall gravity for $\beta>0 $, where $\beta=0$ corresponds to the case in Einstein gravity. Here we choose $M=1$, $a=0.1$, $l=2$ and $n=0$.}
\begin{tabular}{|c|c|c|c|}
\hline
   \multicolumn{1}{|c|}{$\beta$}  &   \multicolumn{1}{c|}{$\omega$}     & \multicolumn{1}{c|}{$\beta$}  &   \multicolumn{1}{c|}{$\omega$} \\
\hline
     0     & 0.319614 - 0.0724516i       &1/10 	  & 0.285076 - 0.0620697i	\\  \hline
    1/82   & 0.317638 - 0.0717914i	   &11/102	  & 0.279023 - 0.060398i	\\  \hline
    1/42    & 0.315405 - 0.0710635i	   & 3/26 	  & 0.272206 - 0.0585497i	\\  \hline
    3/86 	 & 0.312899 - 0.0702637i	   &13/106	  & 0.264487 - 0.0564985i	\\  \hline
    1/22 	 & 0.310097 - 0.0693869i	   &7/54 	  & 0.255696 - 0.0542105i	\\  \hline
   1/18    &0.306972 - 0.0684269i	    &3/22 	  & 0.245607 - 0.0516404i   \\   \hline
    3/46 	 & 0.30349 - 0.0673766i	    &1/7  	  & 0.233909 - 0.0487283i	\\  \hline
    7/94 	 & 0.299613 - 0.066227i	    & 17/114 & 0.220181 - 0.0454018i	\\   \hline
    1/12 	 & 0.295292 - 0.064968i	    &9/58    & 0.203778 - 0.0415378i	\\  \hline
    9/98 	 & 0.29047 - 0.0635871i        &19/118  &0.183683 - 0.0369566i	\\  \hline
		
  \end{tabular}%
  \label{tab:addlabel}%
\end{table}%

\begin{table}[htbp]\small
\centering
\caption{The quasinormal mode frequencies of the gravitational perturbation in a Schwarzschild black hole surrounded by a cloud of strings in Rastall gravity for $\beta>0 $, where $\beta=0$ corresponds to the case in Einstein gravity. Here we choose $M=1$, $l=2$ and $n=0$.}
\begin{tabular}{|c|c|c|c|c|c|}
\hline
  \multicolumn{1}{|c|}{$\beta$}   &   \multicolumn{1}{c|}{$a$}  &   \multicolumn{1}{c|}{$\omega$}   & \multicolumn{1}{c|}{$\beta$}  & \multicolumn{1}{c|}{$a$}  &   \multicolumn{1}{c|}{$\omega$} \\
\hline
  \multirow{23}[0]{*}{0}   & 0     & 0.373609 - 0.0889758i &        \multirow{23}[0]{*}{ 1/10}    & 0     & 0.373609 - 0.0889758i  \\  \cline{2-3} \cline{5-6}
&0.01 &	0.368085 - 0.087252i	 &	&	0.01 &	0.365032 - 0.0862744i		 \\    \cline{2-3} \cline{5-6}
&0.02	&    0.362589 - 0.0855446i	 &	&	0.02	 &    0.356404 - 0.0835761i	 \\    \cline{2-3} \cline{5-6}	
&0.04 &	0.351676 - 0.0821788i	 &	&	0.04	 &    0.338996 - 0.0781593i	\\    \cline{2-3} \cline{5-6}
&0.06 &	0.340873 - 0.0788784i	 &	&	0.06 &	0.321322 - 0.07279i		 \\    \cline{2-3} \cline{5-6}
&0.08 &	0.330197 - 0.0756167i	 &	&	0.08 &	0.303362 - 0.0674277i		 \\    \cline{2-3} \cline{5-6}
&0.10 &  	0.319614 - 0.0724516i	 &	&	0.10	&    0.285076 - 0.0620697i		 \\    \cline{2-3} \cline{5-6}
&0.12 &	0.309144 - 0.0693517i	 &	&	0.12	 &   0.266409 - 0.0567121i		 \\    \cline{2-3} \cline{5-6}
&0.14 &	0.298786 - 0.0663169i	 &	&	0.14 &	0.247293 - 0.0513512i		 \\    \cline{2-3} \cline{5-6}
&0.16 &	0.288541 - 0.0633437i	 &	&	0.16 &	0.227636 - 0.0459809i		 \\    \cline{2-3} \cline{5-6}
&0.18 &	0.278418 - 0.0604432i	 &	&	0.18	&    0.207317 - 0.0405949i		 \\    \cline{2-3} \cline{5-6}
&0.20 & 	0.26841 - 0.0576046i	 &	&	0.20 &	     0.186152 - 0.0351783i		 \\    \cline{2-3} \cline{5-6}
&0.22 &	0.258521 - 0.0548316i	 &	&	0.22 &	0.163867 - 0.0297108i		 \\    \cline{2-3} \cline{5-6}
&0.24 &	0.248753 - 0.0521208i	 &	&	0.24 &	0.140009 - 0.0241565i		 \\    \cline{2-3} \cline{5-6}
&0.26 &	0.239108 - 0.0494809i	 &	&	0.26	&    0.11372 - 0.0184461i		 \\    \cline{2-3} \cline{5-6}
&0.27	&    0.234332 - 0.0481851i	 &	&	0.27 &	0.0991279 - 0.015487i		 \\    \cline{2-3} \cline{5-6}
&0.28 &	0.229587 - 0.0469078i	 &	&	0.28 &	0.0829569 - 0.0124022i	 \\    \cline{2-3} \cline{5-6}	
&0.286 &   0.226755 - 0.0461482i&	&	0.286	 &   0.072082 - 0.0104525i		 \\    \cline{2-3} \cline{5-6}
&0.29	 &   0.224874 - 0.0456422i	 &	&	0.29 &	0.064083 - 0.00908549i	 \\    \cline{2-3} \cline{5-6}	
&0.293&   0.223467 - 0.045268i	 &	&	0.293 &	0.0575201 - 0.00800739i	 \\    \cline{2-3} \cline{5-6}	
&0.296 &	0.222061 - 0.0448953i	 &	&	0.296 &	0.0502678 - 0.00686153i	 \\    \cline{2-3} \cline{5-6}	
&0.298	 & 0.221127 - 0.0446427i &  &	0.298 &	0.0448894 - 0.00604251i	 \\    \cline{2-3} \cline{5-6}	
&0.30  &	0.220192 - 0.0443971i	 &	&	0.30 & 	0.0388694 - 0.00515565i	 \\    \cline{2-3} \cline{5-6}	

\hline
  \end{tabular}%
  \label{tab:addlabel}%
\end{table}%

\begin{table}[htbp]\footnotesize
\centering
\caption{The quasinormal mode frequencies of the gravitational perturbation in a Schwarzschild black hole surrounded by a cloud of strings in Rastall gravity for $\beta=-1/18$ and $-1/8$, where $\beta=0$ corresponds to the case in Einstein gravity. Here we choose $M=1$,  $l=2$ and $n=0$.}
\begin{tabular}{|c|c|c|c|}
\hline
   \multirow{2}[0]{*}{ \diagbox{$a$}{$\beta$}  } & \multicolumn{1}{c|}{\multirow{2}[0]{*}{$0$}} & \multicolumn{1}{c|}{\multirow{2}[0]{*}{$-1/18$}} & \multicolumn{1}{c|}{\multirow{2}[0]{*}{$-1/8$}} \\  
                & \multicolumn{1}{c|}{}  & \multicolumn{1}{c|}{}  & \multicolumn{1}{c|}{}  \\
\hline
0	&0.373609 - 0.0889758i&			0.373609 - 0.0889758i	&		0.373609 - 0.0889758i		\\
0.01	&0.368085 - 0.087252i	&		     0.368578 - 0.0874442i	&		0.368677 - 0.0875301i		\\
0.02	&0.362589 - 0.0855446i&			0.363588 - 0.0859314i	&		0.363804 - 0.0861071i		\\
0.04	&0.351676 - 0.0821788i&			0.353733 - 0.0829618i	&		0.354236 - 0.0833286i		\\
0.06	&0.340873 - 0.0788784i&			0.344042 - 0.0800661i	&		0.3449 - 0.0806377i		\\
0.08	&0.330197 - 0.0756167i&			0.334516 - 0.0772435i	&		0.335794 - 0.0780322i		\\
0.10 	&0.319614 - 0.0724516i&			0.325172 - 0.0744666i	&		0.326928 - 0.0754804i		\\
0.12	&0.309144 - 0.0693517i&			0.315973 - 0.0717906i	&		0.318266 - 0.0730415i		\\
0.14	&0.298786 - 0.0663169i&			0.306938 - 0.0691845i	&		0.30982 - 0.0706808i		\\
0.16	&0.288541 - 0.0633437i&			0.298065 - 0.0666473i	&		0.301585 - 0.0683961i		\\
0.18	&0.278418 - 0.0604432i&			0.289355 - 0.0641782i	&		0.293558 - 0.0661851i		\\
0.20	&0.26841 - 0.0576046i	&		     0.280806 - 0.0617762i	&		0.285735 - 0.0640459i		\\
0.22	&0.258521 - 0.0548316i&			0.272419 - 0.0594402i	&		0.278112 - 0.0619763i		\\
0.24	&0.248753 - 0.0521208i&			0.264193 - 0.0571697i	&		0.270685 - 0.0599744i		\\
0.26	&0.239108 - 0.0494809i&			0.256128 - 0.0549635i	&		0.263449 - 0.0580381i		\\
0.28	&0.229587 - 0.0469078i&			0.248223 - 0.0528188i	&		0.256405 - 0.056168i		\\
0.30 	&0.220192 - 0.0443971i&			0.240475 - 0.0507402i	&		0.249538 - 0.0543547i		\\
0.32	&0.210927 - 0.041954i	&		     0.232887 - 0.0487212i	&		0.242855 - 0.052604i		\\
0.34	&0.201786 - 0.0395782i&			0.225457 - 0.0467628i	&		0.236348 - 0.0509114i		\\
0.36	&0.192783 - 0.0372704i&			0.218185 - 0.0448634i	&		0.230014 - 0.0492753i		\\
0.38	&0.183911 - 0.0350269i&			0.211069 - 0.0430218i	&		0.223848 - 0.047693i		\\
0.40 	&0.175176 - 0.0328518i&			0.204104 - 0.0412383i&		0.217847 - 0.0461635i		\\
0.44	&0.158125 - 0.0287039i&			0.19065 - 0.0378432i	&		0.206328 - 0.043261i		\\
0.48	&0.141648 - 0.0248258i&			0.177803 - 0.0346644i	&		0.195424 - 0.0405495i		\\
0.52	&0.125769 - 0.0212199i&			0.165563 - 0.0316969i	&		0.185109 - 0.0380178i		\\
0.56	&0.110513 - 0.0178881i&			0.153919 - 0.0289317i	&		0.175352 - 0.0356547i		\\
0.60	&0.0959109 - 0.0148325i&			0.142863 - 0.0263604i&		0.166135 - 0.033453i		\\
0.64	&0.0819965 - 0.012055i&			0.132383 - 0.0239746i	&		0.157419 - 0.031397i		\\
0.68	&0.0688098 - 0.00955797i&			0.12247 - 0.0217662i	&		0.14919 - 0.029482i		\\
0.72	&0.0563983 - 0.00734378i&			0.113112 - 0.0197268i	&		0.141417 - 0.0276933i		\\
0.76	&0.0448199 - 0.00541499i&			0.104296 - 0.0178479i	&		0.13408 - 0.0260266i		\\
0.80 	&0.0341463 - 0.00377434i&			0.0960096 - 0.0161209i&		0.127155 - 0.0244728i		\\
0.84	&0.0244707 - 0.00242471i&			0.0882376 - 0.0145377i&		0.120618 - 0.0230226i		\\
0.88	&0.0159194 - 0.00136915i&			0.0809652 - 0.0130901i&		0.114452 - 0.0216708i		\\
0.92	&0.00867958 - 0.000610903i& 		0.0741765 - 0.0117695i&		0.108634 - 0.0204093i		\\
0.96	&0.00307387 - 0.000153336i&		0.0678542 - 0.0105679i&		0.103145 - 0.0192326i		\\

\hline
    \end{tabular}%
  \label{tab:addlabel}%
\end{table}%

\begin{table}[htbp]\footnotesize
\centering
\caption{The quasinormal mode frequencies of the gravitational perturbation in a Schwarzschild black hole surrounded by a cloud of strings in Rastall gravity for $\beta=-3/14$, $-1/3$ and $-19/42$. Here we choose $M=1$,  
  $l=2$ and $n=0$.}
\begin{tabular}{|c|c|c|c|}
\hline
   \multirow{2}[0]{*}{ \diagbox{$a$}{$\beta$}  } & \multicolumn{1}{c|}{\multirow{2}[0]{*}{$-3/14$}} & \multicolumn{1}{c|}{\multirow{2}[0]{*}{$-1/3$}} & \multicolumn{1}{c|}{\multirow{2}[0]{*}{$-19/42$}} \\  
                & \multicolumn{1}{c|}{}  & \multicolumn{1}{c|}{}  & \multicolumn{1}{c|}{}  \\
\hline

0	&0.373609 - 0.0889758i			&0.373609 - 0.0889758i			&0.373609 - 0.0889758i	\\	
0.01	&0.368131 - 0.0874521i			&0.365483 - 0.0868656i			&0.348588 - 0.0828812i	\\	
0.02	&0.362754 - 0.0859613i			&0.357639 - 0.0848351i			&0.326575 - 0.0775351i	\\	
0.04	&0.352291 - 0.083076i			     &0.342745 - 0.0809981i			&0.289673 - 0.0686059i	\\	
0.06	&0.342207 - 0.0803135i			&0.328832 - 0.0774357i			&0.259995 - 0.0614554i	\\	
0.08	&0.332485 - 0.0776681i			&0.315818 - 0.0741223i			&0.23564 - 0.0556078i		\\
0.10 &0.323111 - 0.0751342i			&0.303624 - 0.071035i			     &0.215315 - 0.0507418i	\\	
0.12	&0.314089 - 0.0726779i			&0.292184 - 0.0681535i			&0.198109 - 0.0466327i	\\	
0.14	&0.305372 - 0.0703538i			&0.281436 - 0.0654598i			&0.183365 - 0.0431191i	\\	
0.16	&0.296963 - 0.0681257i			&0.27134 - 0.0629128i			     &0.170598 - 0.0400819i	\\	
0.18	&0.288851 - 0.0659893i			&0.261816 - 0.0605496i			&0.159439 - 0.0374317i	\\	
0.20	&0.281024 - 0.0639402i			&0.252833 - 0.0583303i			&0.149607 - 0.0350999i	\\	
0.22	&0.273471 - 0.0619742i			&0.244351 - 0.0562432i			&0.140882 - 0.033033i		\\
0.24	&0.266181 - 0.0600876i			&0.236333 - 0.054278i			     &0.133088 - 0.0311889i	\\	
0.26	&0.259143 - 0.0582766i			&0.228744 - 0.052425i			     &0.126087 - 0.0295339i	\\	
0.28	&0.252348 - 0.0565377i			&0.221554 - 0.0506757i			&0.119764 - 0.0280406i	\\	
0.30 	&0.245786 - 0.0548676i			&0.214735 - 0.0490223i			&0.114026 - 0.0266867i	\\	
0.32	&0.239448 - 0.0532631i			&0.208262 - 0.0474578i			&0.108798 - 0.0254539i	\\	
0.34	&0.233326 - 0.0517212i			&0.20211 - 0.0459757i			     &0.104014 - 0.0243267i	\\	
0.36	&0.22741 - 0.050239i			     &0.196258 - 0.0445701i			&0.0996215 - 0.0232922i	\\	
0.38	&0.221694 - 0.0488138i			&0.190686 - 0.0432357i			&0.0955744 - 0.0223397i	\\	
0.40 &0.216167 - 0.0474396i			&0.185377 - 0.0419677i			&0.091834 - 0.0214599i	\\	
0.44	&0.205661 - 0.0448521i			&0.17548 - 0.0396132i			      &0.0851453 - 0.0198876i	\\	
0.48	&0.195834 - 0.0424548i			&0.166448 - 0.0374751i			&0.0793404 - 0.0185243i	\\	
0.52	&0.186637 - 0.0402334i			&0.15818 - 0.0355267i			      &0.0742566 - 0.0173312i	\\	
0.56	&0.178021 - 0.0381676i			&0.150589 - 0.0337455i			&0.0697688 - 0.0162787i	\\	
0.60	&0.169942 - 0.0362462i			&0.143601 - 0.0321122i			&0.065779 - 0.0153435i	\\	
0.64	&0.162362 - 0.0344581i			&0.137151 - 0.0306103i			&0.0622093 - 0.0145072i	\\	
0.68	&0.155245 - 0.0327927i			&0.131183 - 0.0292253i			&0.0589975 - 0.0137551i	\\	
0.72	&0.148556 - 0.0312361i			&0.125648 - 0.027945i			      &0.0560927 - 0.0130752i	\\	
0.76	&0.142263 - 0.0297872i			&0.120504 - 0.0267587i			&0.0534534 - 0.0124577i	\\	
0.80 	&0.136343 - 0.0284285i			&0.115714 - 0.0256569i			&0.051045 - 0.0118944i	\\	
0.84	&0.130762 - 0.0271592i			&0.111243 - 0.0246315i			&0.0488389 - 0.0113786i	\\	
0.88	&0.1255 - 0.0259672i			     &0.107063 - 0.0236752i			&0.0468108 - 0.0109045i	\\	
0.92	&0.12054 - 0.02485i			     &0.103147 - 0.0227786i			&0.0449402 - 0.0104674i	\\	
0.96	&0.115852 - 0.023803i			     &0.0994759 - 0.0219452i		     &0.0432096 - 0.0100631i	\\	

\hline
    \end{tabular}%
  \label{tab:addlabel}%
\end{table}%

In Fig. 4.1 and Fig. 4.2,  we draw the graphs of real parts and negative imaginary parts of quasinormal mode frequencies with respect to the parameter  $\beta $ for $a=0.1$ when $\beta >0$. From the two figures, we can see that both the real part and the absolute value of the imaginary part  decrease with the increase of $\beta$. In Fig. 4.3 and Fig. 4.4, we draw the graphs of real parts and negative imaginary parts of quasinormal mode frequencies with respect to the parameter $a$ for the fixed $\beta=1/10 $, where the black curves represent the real part and the negative imaginary part in Einstein gravity. From the two figures, we can see that, on the one hand, both the real part and the absolute value of the imaginary part decrease with the increase of $a$; and on the other hand, the real part and the absolute value of the imaginary part in Rastall gravity is significantly lower than that in Einstein gravity at the same value of $a$. For example, when the parameter $a=0.2$, the red curve representing Rastall gravity is significantly lower than the black curve representing Einstein gravity in Fig. 4.3 and in Fig. 4.4.

\begin{figure}
\centering
\begin{minipage}[t]{0.6\linewidth}
\includegraphics[width=105mm]{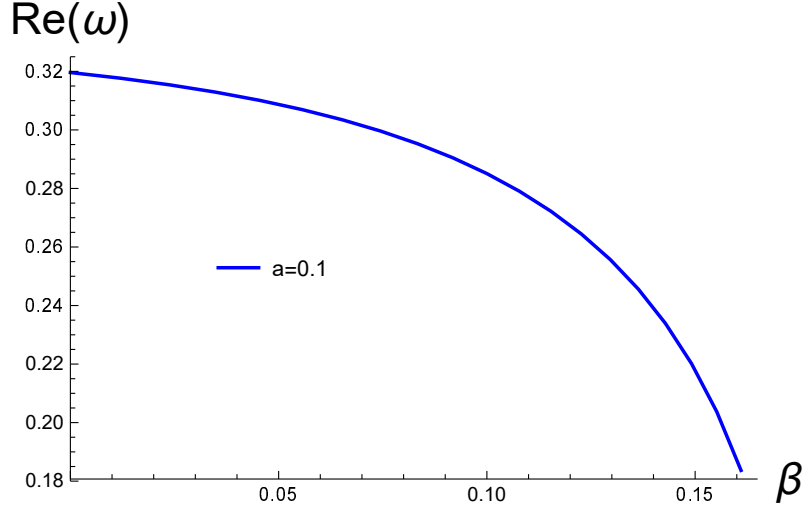}
\caption*{Fig. 4.1 Graph of real parts of quasinormal mode frequencies with respect to the parameter $\beta $. Here we choose $M=1$,  $a=0.1$, $l = 2$ and $n=0$.}
\label{fig41} 
\end{minipage}
\end{figure}

\begin{figure}
\centering
\begin{minipage}[t]{0.6\linewidth}
\includegraphics[width=105mm]{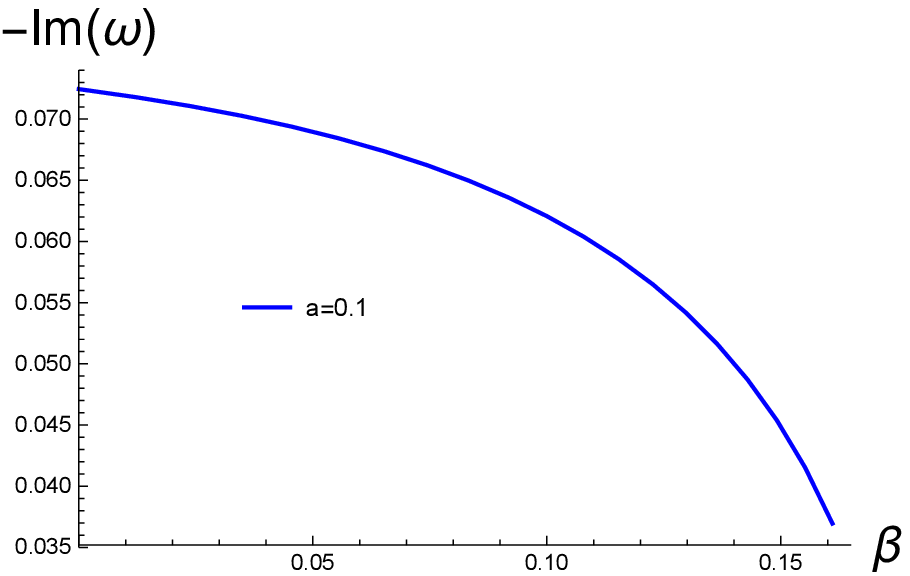}
\caption*{Fig. 4.2 Graph of negative imaginary parts of quasinormal mode frequencies with respect to the parameter $\beta $. Here we choose $M=1$, $a=0.1$, $l = 2$ and $n=0$.}
\label{fig42} 
\end{minipage}
\end{figure}

\begin{figure}
\centering
\begin{minipage}[t]{0.6\linewidth}
\includegraphics[width=105mm]{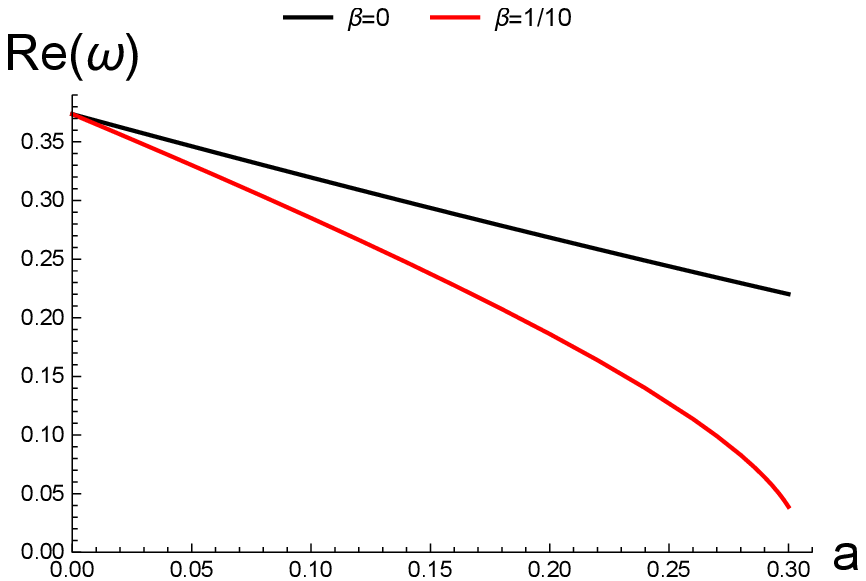}
\caption*{Fig. 4.3 Graph of real parts of quasinormal mode frequencies with respect to the parameter $a $. Here we choose $M=1$, $l = 2$ and $n=0$.}
\label{fig41} 
\end{minipage}
\end{figure}

\begin{figure}
\centering
\begin{minipage}[t]{0.6\linewidth}
\includegraphics[width=105mm]{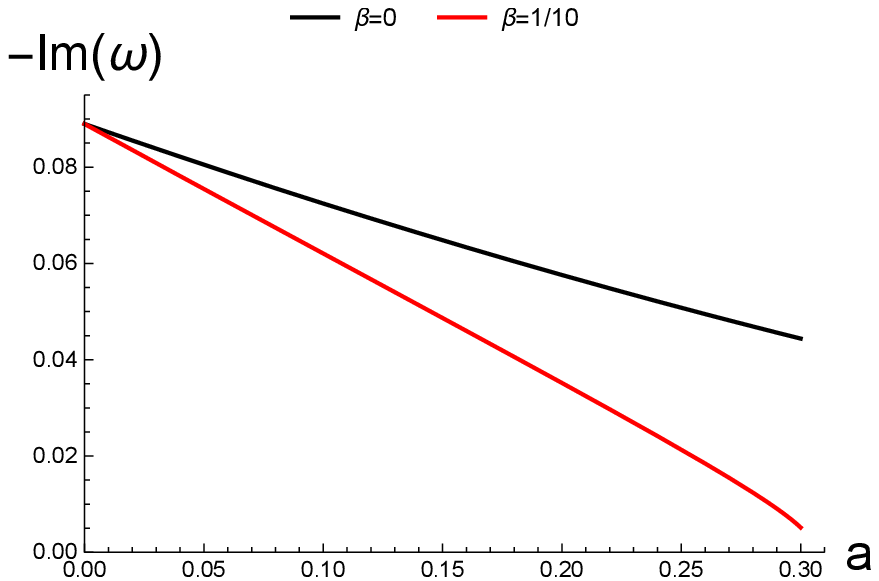}
\caption*{Fig. 4.4 Graph of negative imaginary parts of quasinormal mode frequencies with respect to the parameter $a $. Here we choose $M=1$, $l = 2$ and $n=0$.}
\label{fig42} 
\end{minipage}
\end{figure}

In Fig. 4.5 and Fig. 4.6, we draw the graphs of real parts and negative imaginary parts of quasinormal mode frequencies with respect to the parameter $a$ for different values of  $\beta $ when $\beta <0$, where the black curves represent the real part and the negative imaginary part in Einstein gravity. From the two figures, we can see that, on the one hand, both the real part and the absolute value of the imaginary part decrease with the increase of $a$; and on the other hand, when $\beta$ is small, e.g. $\beta=-19/42$ (the purple curve), both the real part and the absolute value of the imaginary part decrease fast at a small $a$, e.g. $a<0.2$, and then decrease very slowly at a large $a$, e.g. $a > 0.8$, which makes the real part and the absolute value of the imaginary part in Rastall gravity larger than that in  Einstein gravity as $a$ approaches 1. We also find that for the same value of $a$, both the real part and the absolute value of the imaginary part increase at first and then decrease with the decrease of $\beta$. In addition, we notice that for the same parameter $a$, the real part and the absolute value of the imaginary part of quasinormal mode frequencies in Rastall gravity are greater or smaller than that in Einstein gravity, depending on the value of the parameter $\beta$. For example, in Fig. 4.6, when $a=0.4$, we can easily see that the absolute value of the imaginary part of quasinormal mode frequencies in Rastall gravity (the red curve corresponding to $\beta=-1/18$) is larger than that of Einstein gravity (the black curve corresponding to $\beta=0$), while the absolute value of the imaginary part of quasinormal mode frequencies in Rastall gravity (the purple curve corresponding to $\beta=-19/42$) is smaller than that of Einstein gravity (the black curve corresponding to $\beta=0$).

\begin{figure}
\centering
\begin{minipage}[t]{0.6\linewidth}
\includegraphics[width=105mm]{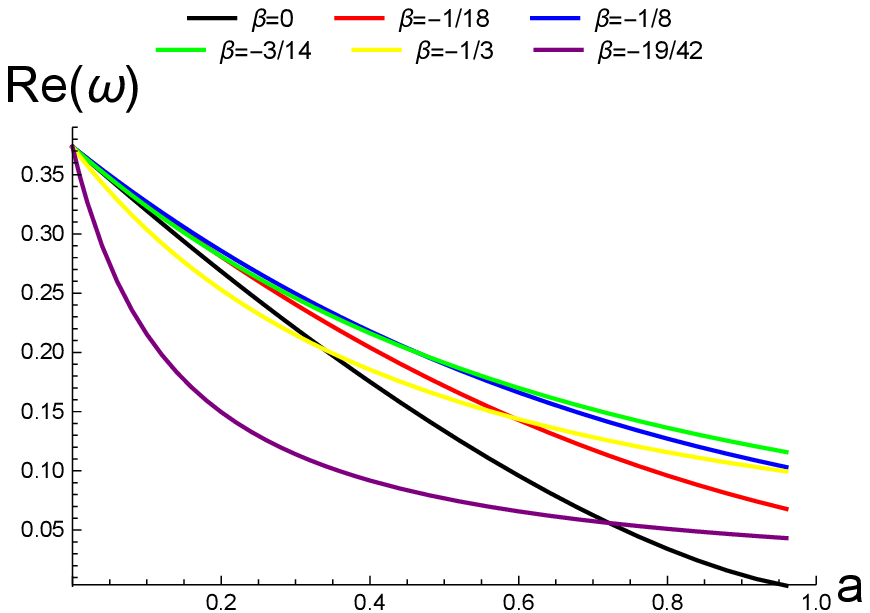}
\caption*{Fig. 4.5 Graph of real parts of quasinormal  mode frequencies with respect to the parameter  $a$ for different values of $\beta $. Here we choose $M=1$, $l = 2$ and $n=0$.}
\label{fig43} 
\end{minipage}
\end{figure}

\begin{figure}
\centering
\begin{minipage}[t]{0.6\linewidth}
\includegraphics[width=105mm]{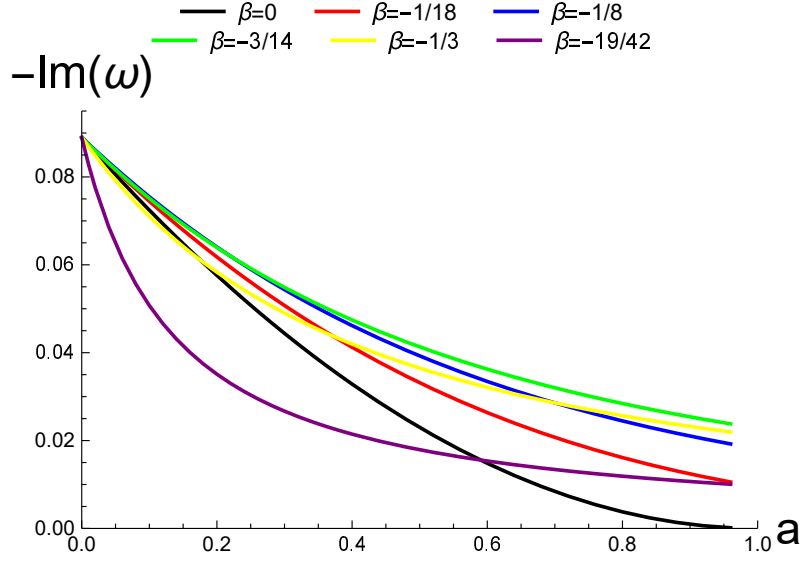}
\caption*{Fig. 4.6 Graph of negative imaginary parts of quasinormal mode frequencies with respect to the parameter $a$ for different values of $\beta $. Here we choose $M=1$, $l = 2$ and $n=0$.}
\label{fig44} 
\end{minipage}
\end{figure}

\subsection{Quasinormal mode frequencies  at the eikonal limit calculated  via the unstable null geodesics  } 

As for the method to calculate the quasinormal mode frequencies at the eikonal limit via the unstable null geodesics of black holes, it was first proposed~\cite{PVC} by Cardoso et al. for the case of the static spherically symmetric spacetime. In the eikonal limit ($l\gg 1$), the scalar, electromagnetic and gravitational perturbations  in the background of the static spherically symmetric metric, see Eq.~(\ref{5}), have the same behavior~\cite{WLX}, and the formula for calculating the quasinormal mode frequencies $\omega _{l\gg 1}$ through the unstable null geodesics reads~\cite{PVC}: 
\begin{equation}
\label{101}
\omega _{l\gg 1}=l\Omega -i(n+\frac{1}{2})|\lambda_{\rm L}|,
\end{equation}
where  $\Omega$ and $\lambda_{\rm L}$ are the angular velocity and the Lyapunov exponent at the unstable null geodesics, and they determine the real part and the imaginary part of  the quasinormal mode frequencies, respectively.
Furthermore, the two quantities can be expressed~\cite{PVC, WLX} as follows:
\begin{equation}
\label{102}
\Omega =\frac{\sqrt{f(r_{c})}}{r_{c}},        ~\ ~\ ~\ ~\ ~\   \lambda_{\rm L} =\sqrt{\frac{f(r_{c})(2f(r_{c})-r_{c}^{2}{f}''(r_{c}))}{2r_{c}^{2}}},
\end{equation}
where $r_{c}$ represents the radius of the unstable null geodesics and is determined by the following relation,
\begin{equation}
\label{1021}
2f(r_{c})-r_{c}\left.\frac{\mathrm{d} f(r)}{\mathrm{d} r}\right|_{r=r_{c}}=0.
\end{equation}
By substituting Eq.~(\ref{30}) into Eq.~(\ref{102}), we draw the graphs of $\Omega$ and $\lambda_{\rm L}$ with respect to $\beta$ and $a$, respectively.

In Fig. 4.7 and Fig. 4.8,  we draw the angular velocity  $\Omega$ and the Lyapunov exponent $\lambda_{\rm L}$ with respect to the parameter  $\beta $ for $a=0.1$  when $\beta >0$. From the two figures, we can see that both  $\Omega$ and  $\lambda_{\rm L}$  decrease with the increase of $\beta$. 

In Fig. 4.9 and Fig. 4.10,  we draw the angular velocity  $\Omega$ and the Lyapunov exponent $\lambda_{\rm L}$  with respect to the parameter  $a$ for $\beta =0.1$,  where the black curves represent the angular velocity  $\Omega$ and the Lyapunov exponent $\lambda_{\rm L}$  in Einstein gravity. From the two figures, we can see that both  $\Omega$ and  $\lambda_{\rm L}$  decrease with the increase of $a$. 

In Fig. 4.11 and Fig. 4.12, we draw the graphs of the angular velocity  $\Omega$ and the Lyapunov exponent $\lambda_{\rm L}$  with respect to the parameter $a$ for different values of  $\beta $ when $\beta <0$,  where the black curve represents the case of Einstein gravity. From the two figures, we can see that both  $\Omega$ and  $\lambda_{\rm L}$  decrease with the increase of $a$.

By comparing Figs. 4.1-4.6 with Figs. 4.7-4.12, we can easily see that Figs. 4.1-4.6 are very similar to Figs. 4.7-4.12, which  shows that  the conclusions made by the method of Regge and Wheeler together with the high order WKB-Pad\'e approximation are identical with those by  the unstable null geodesics at the eikonal limit. Finally, we have the following results through the comparison and analysis of Figs. 4.1-4.12: 
\begin{itemize}
\item The increase of the parameter $a$ in Rastall gravity makes the gravitational wave decay slowly.
\item For the same parameter $a$, when $\beta>0$, the gravitational waves in Rastall gravity always decay more slowly  than those in Einstein  gravity, but when  $\beta<0$, the gravitational waves in Rastall gravity decay more slowly or faster than those in Einstein  gravity, which depends on the value of the parameter $\beta$.
\item For $\beta<0$, the gravitational waves in Rastall gravity decay faster than those in Einstein gravity as $a$ approaches 1.
\end{itemize}

\begin{figure}
\centering
\begin{minipage}[t]{0.6\linewidth}
\includegraphics[width=105mm]{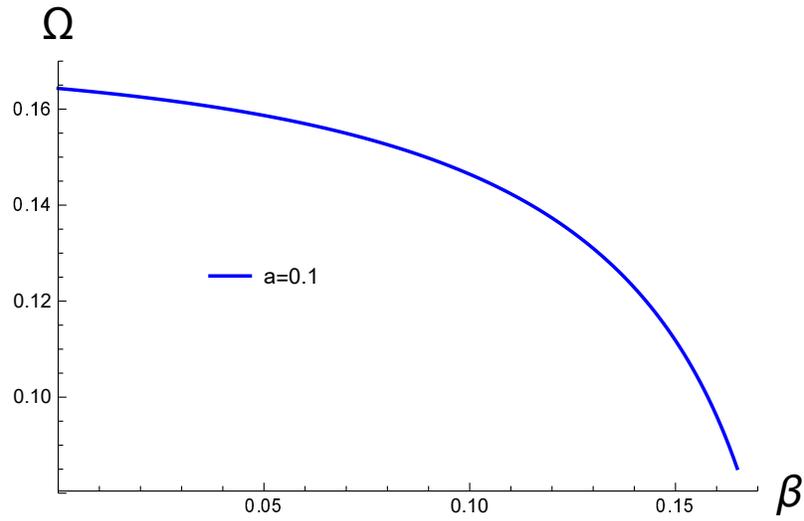}
\caption*{Fig. 4.7 Graph of the angular velocity  $\Omega$   with respect to the parameter $\beta $. Here we choose $M=1$ and $a=0.1$.}
\label{fig43} 
\end{minipage}
\end{figure}

\begin{figure}
\centering
\begin{minipage}[t]{0.6\linewidth}
\includegraphics[width=105mm]{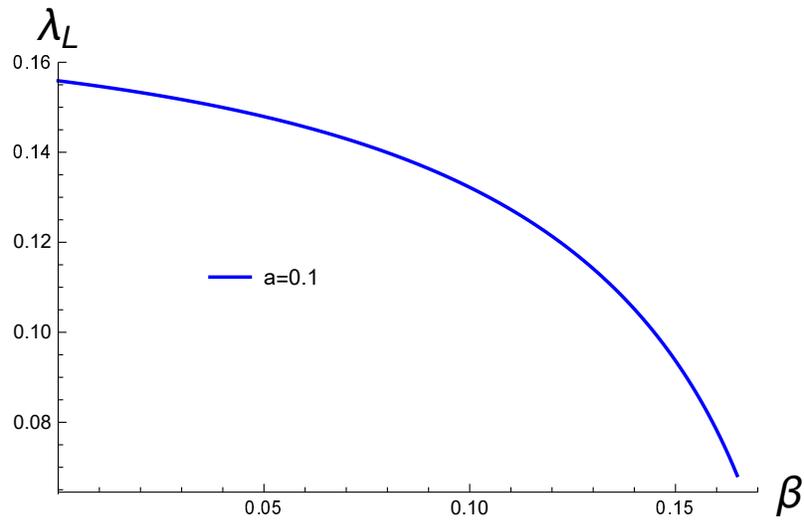}
\caption*{Fig. 4.8 Graph of the Lyapunov exponent $\lambda_{\rm L}$   with respect to the parameter $\beta $. Here we choose $M=1$ and $a=0.1$.}
\label{fig43} 
\end{minipage}
\end{figure}

\begin{figure}
\centering
\begin{minipage}[t]{0.6\linewidth}
\includegraphics[width=105mm]{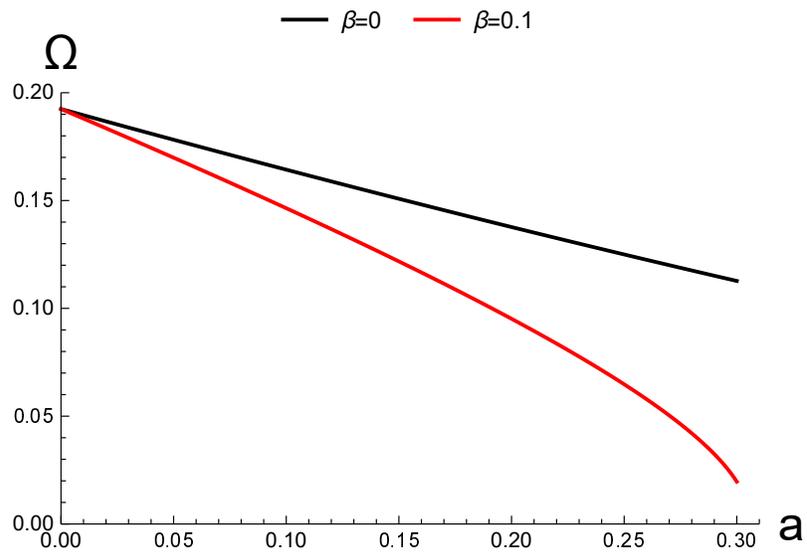}
\caption*{Fig. 4.9 Graph of the angular velocity  $\Omega$  with respect to the parameter  $a$. Here we choose $M=1$.}
\label{fig43} 
\end{minipage}
\end{figure}

\begin{figure}
\centering
\begin{minipage}[t]{0.6\linewidth}
\includegraphics[width=105mm]{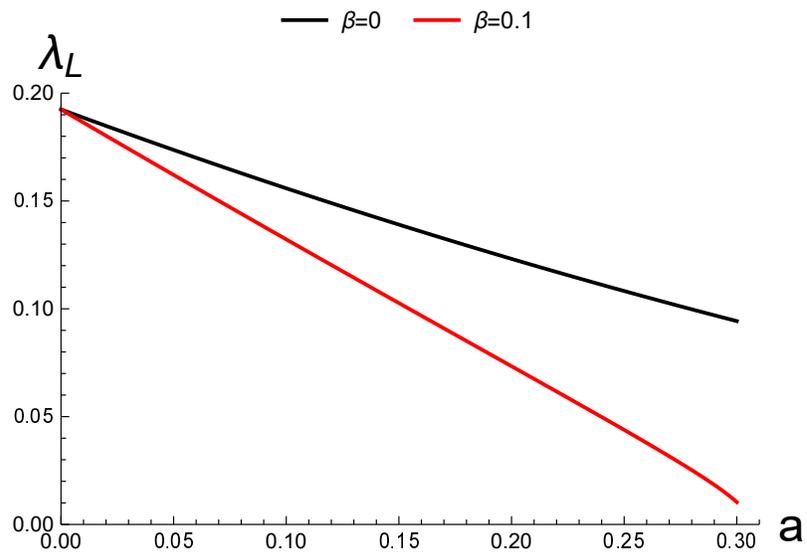}
\caption*{Fig. 4.10 Graph of the Lyapunov exponent $\lambda_{\rm L}$  with respect to the parameter  $a$. Here we choose $M=1$.}
\label{fig43} 
\end{minipage}
\end{figure}

\begin{figure}
\centering
\begin{minipage}[t]{0.6\linewidth}
\includegraphics[width=105mm]{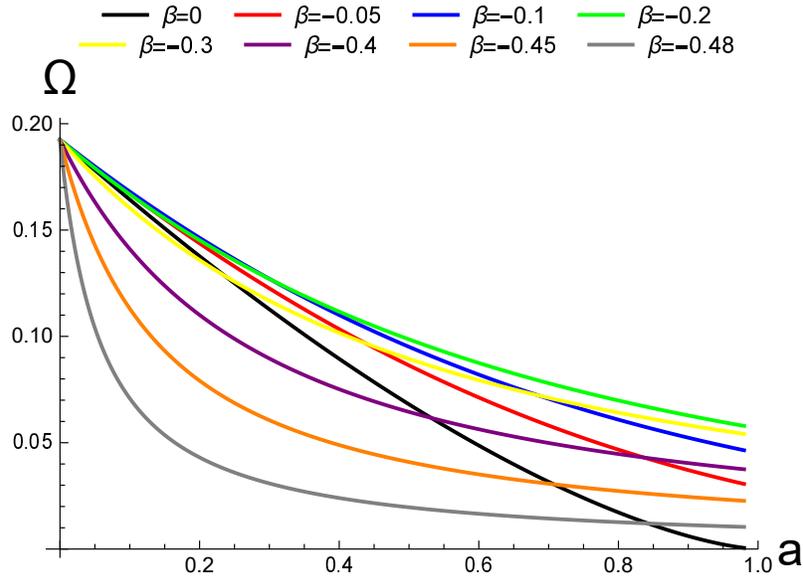}
\caption*{Fig. 4.11 Graph of the angular velocity  $\Omega$  with respect to the parameter  $a$ for different values of $\beta $. Here we choose $M=1$.}
\label{fig43} 
\end{minipage}
\end{figure}

\begin{figure}
\centering
\begin{minipage}[t]{0.6\linewidth}
\includegraphics[width=105mm]{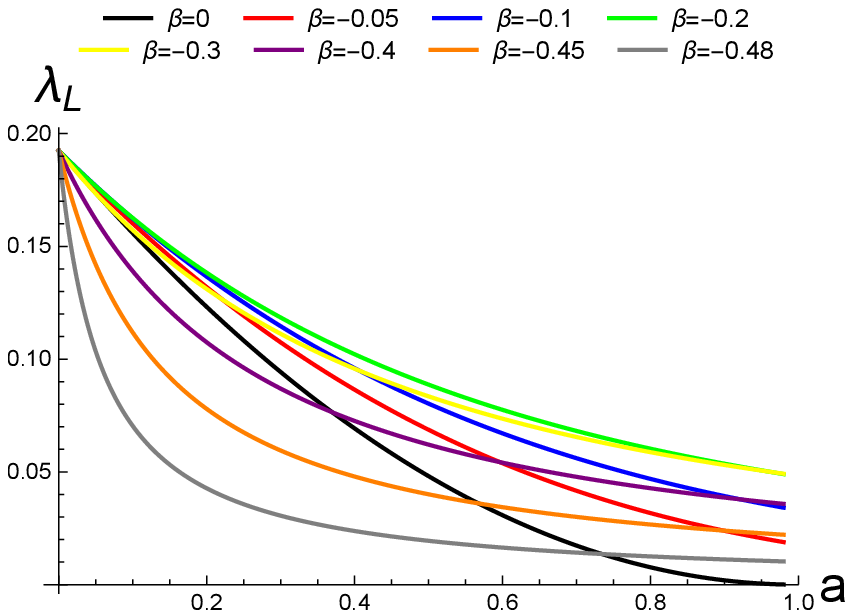}
\caption*{Fig. 4.12 Graph of the Lyapunov exponent $\lambda_{\rm L}$  with respect to the parameter  $a$ for different values of $\beta $. Here we choose $M=1$.}
\label{fig43} 
\end{minipage}
\end{figure}

\section{Area and entropy spectra via adiabatic invariance}
In this section, based on the method of adiabatic invariance of black holes~\cite{P30} and its later improvement~\cite{P13} for deriving the quantized entropy spectrum, we study the area spectrum and entropy spectrum of a Schwarzschild black hole surrounded by  a cloud of strings in Rastall gravity. By applying the Wick rotation in Lorentzian time and transforming $t$ to $-i\tau$, we write the Euclideanized form of the metric, 
\begin{equation}
\label{35}
ds^{2}=-f(r)d\tau^{2}-f^{-1}(r)dr^{2}-r^{2}(d\theta^{2}+\sin^{2}\theta d\phi ^{2}),
\end{equation}
where $f(r)$ is given in Eq.~(\ref{30})
 and $\tau$ means the Euclidean time. For a neutral and static spherically symmetric black hole spacetime, the only dynamic freedom of adiabatic invariants is the radial coordinate $r$, so the adiabatic invariant can be expressed~\cite{P31} as
\begin{equation}
\label{36}
\mathcal{L}=\oint p_{r}dq_{r}=\oint\int_{0}^{p_{r}}dp'_{r}dr,
\end{equation}
where $p_{r}$ is the conjugate momentum to the coordinate $q_{r}$. 

It is known that the equation of motion of a massless particle is~\cite{P32} a radial null geodesic, $\dot{r}=\frac{\mathrm{d} r}{\mathrm{d} \tau }$, and the equation of motion of a massive particle is~\cite{P33} the phase velocity, $\dot{r}=v_{p}$. However, if one only considers the outgoing path, one does not care~\cite{P34} whether the particle has mass or not.
According to Hamilton's canonical equation, we have
\begin{equation}
\label{37}
\dot{r}=\frac{\mathrm{d} r}{\mathrm{d} \tau }=\frac{\mathrm{d} H'}{\mathrm{d} p'_{r}},
\end{equation}
where $H'=M'$ and $M'=M-\hbar\omega'$, $M'$ is the mass of the black hole from which a particle with energy $E'=\hbar\omega '$ tunnels through its horizon. By substituting  Eq.~(\ref{37}) into Eq.~(\ref{36}), we obtain the following formula,
\begin{equation}
\label{38}
\mathcal{L}=\oint p_{r}dq_{r}=\oint\int_{0}^{H}\frac{dH'}{\dot{r}}dr=\oint\int_{0}^{M}dM'd\tau.
\end{equation}
Since any background spacetime with the horizon in Kruskal coordinates has periodicity with respect to the Euclidean time, we assume that the particles moving in such a background spacetime  have the same periodicity as the background spacetime, and this period satisfies~\cite{P35} the following relationship,
\begin{equation}
\label{39}
T=\oint d\tau =\frac{2\pi }{\kappa _{r_{\rm H}}}=\frac{\hbar}{T_{\rm{BH}}}.
\end{equation}

Using  Eq.~(\ref{39}), we rewrite Eq.~(\ref{38}) as follows:
\begin{equation}
\label{40}
\mathcal{L}=\hbar\int_{0}^{M}\frac{dM'}{T'_{\rm{BH}}}=\hbar\int_{0}^{r_{\rm H}}\frac{1}{T'_{\rm{BH}}}\frac{dM'}{dr'_{\rm H}}dr'_{\rm H}.
\end{equation}
Next, we deduce the following formula from Eq.~(\ref{32}),
\begin{equation}
\label{41}
\frac{\mathrm{d} M}{\mathrm{d} r_{\rm H}}=\frac{1}{2} \left(1+\frac{a (1-2 \beta ) r_{\rm H}^{\frac{4 \beta }{1-2 \beta }}}{4 \beta -1}\right).
\end{equation}
After substituting Eq.~(\ref{34}) and Eq.~(\ref{41}) into Eq.~(\ref{40}), we finally recast this adiabatic invariant  as follows:
\begin{equation}
\label{42}
\mathcal{L}=\pi r^{2}_{\rm H}=\frac{A}{4},
\end{equation}
where A is the area of the event horizon of the black hole. In addition, we know its quantized form from the Bohr-Sommerfeld quantization rule, $\mathcal{L}=2\pi n\hbar$, $n=0, 1, 2, 3, \dots$, so we have the quantized area of the black hole,
\begin{equation}
\label{43}
A_{n}=8\pi n \hbar.
\end{equation}

For a black hole in Einstein gravity, the entropy $S$ is one quarter of the area of the event horizon, which is also known as the Bekenstein entropy, $S=\frac{A}{4}$. But for the modified gravity, the modified terms other than Einstein tensor usually modify Bekenstein entropy, which has been shown in Refs.~\cite{PMRCS,PAMCRG,PCRCO}. 
Therefore, for Rastall gravity as a modified theory of gravity, the relationship between the horizon entropy and the horizon area is very likely not to satisfy the Bekenstein conjecture.
In fact, the corresponding relationship has been given~\cite{P24} for Rastall gravity by generalizing the Misner-Sharp mass and the relevant first law of thermodynamics,  

\begin{equation}
\label{107}
S=\left(\frac{6\beta -1}{4\beta -1}\right)\frac{A}{4\hbar}.
\end{equation}
Obviously, when $\beta =0$, it turns back to Bekenstein entropy. By substituting Eq.~(\ref{43}) into Eq.~(\ref{107}), we obtain the quantized entropy,
\begin{equation}
\label{108}
S_n=\left(\frac{6\beta -1}{4\beta -1}\right)2\pi n.
\end{equation}

As a result, we reach the area spectrum and the entropy spectrum of the Schwarzschild black hole surrounded by a cloud of strings in Rastall gravity,
\begin{equation}
\label{44}
\Delta A=A_{n}-A_{n-1}=8\pi \hbar,   ~\ ~\ ~\ ~\   \Delta S=S_{n}-S_{n-1}=\left(\frac{6\beta -1}{4\beta -1}\right)2\pi.
\end{equation}

\section{ Area and entropy spectra via gravitational wave periodicity}
In this section, we investigate the area spectrum and entropy spectrum of a Schwarzschild black hole surrounded by a cloud of strings in Rastall gravity by utilizing the method of the periodic property of the outgoing gravitational wave~\cite{P21}. We have two ways to calculate wave functions. One is to substitute~\cite{P36} the assumed form of wave functions, $\Phi =\frac{1}{4\pi \omega ^{{1}/{2}}}\frac{1}{r}R_{\omega }(r,t)Y_{l,m}(\theta ,\phi )$, and Eq.~(\ref{5}) into the Klein-Gordon equation, 
\begin{equation}
\label{45}
g^{\mu \nu }\partial_{\mu }\partial_{\nu}\Phi +\frac{m^{2}}{\hbar^{2}}\Phi =0, 
\end{equation}
and the other is to resort the Hamilton-Jacobi equation,
\begin{equation}
\label{46}
g^{\mu \nu }\partial_{\mu }{\bf S}\partial_{\nu}{\bf S}-m^{2} =0,  
\end{equation}
where the wave function $\Phi$ and  the action $\bf S$ satisfy the following relationship,
\begin{equation}
\label{47}
\Phi=\mathrm{exp}[\frac{i}{\hbar}{\bf S}(t,r,\theta ,\phi )].
\end{equation}

Because the spacetime of the black hole we consider is spherically symmetric, the action $\bf S$ can be decomposed~\cite{P37,P38} as  
\begin{equation}
\label{48}
{\bf S}(t,r,\theta ,\phi )=-Et+W(r)+J(\theta ,\phi ),
\end{equation}
where $E=-\frac{\partial {\bf S}}{\partial t}$ represents  the energy of the emitted particles observed at infinity. Considering $J(\theta,\phi)=0$, $W(r)=i\pi E/f'(r_{\rm H})$ and only the outgoing wave near the outside horizon, one can express the wave function $\Phi$ there as follows,
\begin{equation}
\label{49}
\Phi =\mathrm{exp}\left(-\frac{i}{\hbar}\emph{Et}\right)\Psi(r_{\rm H}),    
\end{equation}
where $\Psi(r_{\rm H})=\mathrm{exp}\left[\frac{-\pi E}{\hbar f'(r_{\rm H})}\right]$ and $f'(r_{\rm H})=\left.\frac{\mathrm{d} f(r)}{\mathrm{d} r}\right|_{r=r_{\rm H}}$. From the above function, we find that $\Phi$ is a periodic wave function with the period, 
\begin{equation}
\label{50}
T=\frac{2\pi \hbar}{E}.
\end{equation}
Considering the relation $E=\hbar \omega$, we obtain
\begin{equation}
\label{51}
T=\frac{2\pi}{\omega}.
\end{equation}
In addition, we know that $T=\frac{\hbar}{T_{\rm{BH}}}$ from Eq.~(\ref{39}), so we derive the following formula,
\begin{equation}
\label{52}
\omega =\frac{2\pi }{T}=\frac{2\pi }{\frac{\hbar}{T_{\rm{BH}}}}=\frac{2\pi T_{\rm{BH}}}{\hbar}.
\end{equation}

According to Hod's idea~\cite{P15}, the change in the area of the event horizon for the Schwarzschild black hole surrounded by a cloud of strings in Rastall gravity is
\begin{equation}
\label{53}
\Delta A=8\pi r_{\rm H}\mathrm{d}r_{\rm H}=8\pi r_{\rm H}\mathrm{d}M\frac{\mathrm{d} r_{\rm H}}{\mathrm{d} M}=8\pi r_{\rm H}\hbar\omega \frac{\mathrm{d} r_{\rm H}}{\mathrm{d} M}.
\end{equation}
Substituting Eq.~(\ref{41}) and Eq.~(\ref{52}) into Eq.~(\ref{53}), we work out
\begin{equation}
\label{54}
\Delta A=8\pi r_{\rm H}\hbar\frac{2\pi T_{\rm{BH}}}{\hbar}\frac{\mathrm{d} r_{\rm H}}{\mathrm{d} M}=16\pi ^{2}r_{\rm H}T_{\rm{BH}}\frac{\mathrm{d} r_{\rm H}}{\mathrm{d} M}= 16\pi ^{2}r_{\rm H} \frac{\hbar }{2 \pi r_{\rm H}}=8\pi\hbar.
\end{equation}
Then using Eq.~(\ref{107}), we deduce the corresponding black hole entropy spectrum,
\begin{equation}
\label{55}
\Delta S=\left(\frac{6\beta -1}{4\beta -1}\right)\frac{\Delta A}{4\hbar}=\left(\frac{6\beta -1}{4\beta -1}\right)2\pi.
\end{equation}

\section{Conclusion }

We first give the solution of the Schwarzschild black hole surrounded by a cloud of strings in Rastall gravity, and then analyze the characteristics of the metric function and the Hawking temperature of this model. We plot the graphs of the metric function $f(r)$ with respect to $r$  and the graphs of the Hawking temperature $T_{\rm {BH}}$ with respect to the event horizon radius $r_{\rm H}$ for different values of $a$ when $\beta >0$ and $\beta <0$, respectively. We find that the increase of the parameter $a$ causes the function $f(r)$ and the Hawking temperature $T_{\rm{BH}}$  to decrease in Rastall gravity. Through the analysis of the black hole metric, we deduce that the solution of the static spherically symmetric black hole surrounded by a cloud of strings in Rastall gravity can be transformed into the solution of the same black hole surrounded by quintessence in Einstein gravity when $\beta>0$, which provides a possibility for a string fluid as a candidate of the dark energy. However, when $\beta<0$, the spacetime of our black hole model is a kind of asymptotically flat spacetime. We use the high order WKB-Pad\'e approximation to calculate the quasinormal mode frequencies of the odd parity gravitational perturbation. Then, we plot the graphs of real parts and negative imaginary parts of quasinormal mode frequencies with respect to the parameter $a$ for $\beta=1/10 $ when $\beta >0$ and the graphs of real parts and negative imaginary parts of quasinormal mode frequencies with respect to the parameter $a$ for different values of $\beta$ when $\beta <0$. We also adopt the unstable null geodesics of the black hole to calculate the quasinormal mode frequencies at the eikonal limit and do the same analysis. The two strategies give rise to the identical conclusion, that is, the increase of the parameter $a$ in Rastall gravity causes the decay of the gravitational wave to slow down. In addition, we utilize the method of the adiabatic invariant integral and the method of the periodic property of outgoing waves to derive the area spectrum and entropy spectrum of the black hole model, respectively. The results show that the area spectrum and entropy spectrum of the Schwarzschild black hole surrounded by a cloud of strings in Rastall gravity are equidistant spaced. The former is same as that of Einstein gravity, while the latter is different, depending on the Rastall parameter $\beta$. It should be noted that the parameter $a$ related to strings does not affect the area spectrum and entropy spectrum of the black hole, which is similar to the phenomenon that the quintessence parameter has no effects~\cite{P31} on the area spectrum and entropy spectrum of the quantum-corrected Schwarzschild black hole surrounded by quintessence.

\section*{Acknowledgments}
This work was supported in part by the National Natural Science Foundation of China under grant No.11675081. The authors would like to thank the anonymous referees for the helpful comments that improve this work greatly.


\end{document}